\documentclass[journal,transmag]{IEEEtran}
\usepackage[top=0.60in, bottom=0.60in, left=0.60in, right=0.60in]{geometry}
\IEEEoverridecommandlockouts
\usepackage{booktabs}
\usepackage{setspace}
\usepackage{multirow}
\usepackage{color}
\usepackage{float}
\usepackage{cite}
\usepackage{enumitem}  
\usepackage[utf8]{inputenc}
\usepackage{lmodern}
\usepackage{tabularx}
\usepackage{amssymb}
\usepackage{csquotes}
\usepackage[justification=centering]{caption}
\usepackage{blindtext}
\usepackage{wrapfig}
\usepackage{graphicx}
\usepackage[bottom]{footmisc}
\DeclareGraphicsExtensions{.eps,.pdf,.gif,.png,.jpg}
\usepackage{amsthm}
\usepackage{array}
\usepackage{algorithm,algorithmic}
\usepackage{booktabs}
\usepackage{graphicx}
\usepackage{epstopdf}\usepackage[utf8]{inputenc}
\usepackage[hyphens,spaces,obeyspaces]{url}
\usepackage[english]{babel}
\usepackage{verbatim} 
\usepackage{lineno,nohyperref}
\usepackage{epstopdf} %converting to PDF
\modulolinenumbers[5]
\newtheorem{theorem}{Theorem}
\newtheorem{definition}{Definition}
\RequirePackage{filecontents}
\usepackage{ragged2e}
\usepackage{url} 
\usepackage[T1]{fontenc}
\usepackage[utf8]{inputenc}
\usepackage{microtype} %remove empy space in URL

\newcommand{\vect}[1]{\boldsymbol{#1}}
\newcommand*{\email}[1]{#1}
\PassOptionsToPackage{hyphens}{url}
\newtheoremstyle{mystyle}%                % Name
{}%                                     % Space above
{}%                                     % Space below
{\itshape}%                                     % Body font
{}%                                     % Indent amount
{\bfseries}%                            % Theorem head font
{.}%                                    % Punctuation after theorem head
{ }%                                    % Space after theorem head, ' ', or \newline
{\thmname{#1}\thmnumber{ #2}\thmnote{ (#3)}}%                                     % Theorem head spec (can be left empty, meaning `normal')

\theoremstyle{mystyle}

\newcounter{subassumption}[asu]

\makeatletter
\renewcommand{\p@subassumption}{\theasu}% Counter prefix.
\makeatother
\usepackage{amsthm}
\usepackage{xpatch,amsthm}
\makeatletter
\xpatchcmd{\@thm}{\fontseries\mddefault\upshape}{}{}{} % same font as thm-header
\makeatother

\makeatother
\usepackage{cite}
\usepackage{amsmath,amssymb,amsfonts}
\usepackage{algorithmic}
\usepackage{graphicx}
\usepackage{textcomp}
\usepackage{xcolor}
\def\BibTeX{{\rm B\kern-.05em{\sc i\kern-.025em b}\kern-.08em
		T\kern-.1667em\lower.7ex\hbox{E}\kern-.125emX}}

\begin{document}
	\title{Two-level Closed Loops for RAN Slice Resources Management Serving Flying  and Ground-based Cars}
	 \author{Anselme~Ndikumana,~
		Kim~Khoa~Nguyen,~\IEEEmembership{Senior~Member,~IEEE,~}\\
		and~Mohamed~Cheriet,~\IEEEmembership{Senior~Member,~IEEE,~}
		\IEEEcompsocitemizethanks{
			\IEEEcompsocthanksitem Anselme Ndikumana, Kim Khoa Nguyen, and Mohamed Cheriet are  with Synchromedia Lab, École de
			Technologie Supérieure, Université du Québec, QC, Canada, E-mail: (\email{anselme.ndikumana.1@ens.tsmtl.ca; kim-khoa.nguyen@etsmtl.ca; Mohamed.Cheriet@etsmtl.ca}).
			%\IEEEcompsocthanksitem 	``The authors thank Mitacs, Ciena, and ENCQOR for funding this research under the IT13947 grant''.
		}}
	\maketitle
	\begin{abstract}
		Flying and ground-based cars require various services such as autonomous driving, remote pilot, infotainment, and remote diagnosis. Each service requires specific Quality of Service (QoS) and network features. Therefore, network slicing can be a solution to fulfill the requirements of various services. Some services, such as infotainment, may have similar requirements to serve flying and ground-based cars. Therefore, some slices can serve both kinds of cars. However, when network slice resource sharing is too aggressive, slices can not meet QoS requirements, where resource under-provisioning causes the violation of QoS, and resource over-provisioning causes resources under-utilization. We propose two closed loops for managing RAN slice resources for cars to address these challenges. First, we present an auction mechanism for allocating  Resource Block (RB) to the tenants who provide services to the cars using slices. Second, we design one closed loop that maps slices and services of tenants to virtual Open Distributed Units (vO-DUs) and assigns RB to vO-DUs for management purposes. Third, we design another closed loop for intra-slices RB scheduling to serve cars. Fourth, we present a reward function that interconnects these two closed loops to satisfy the time-varying demands of cars at each slice while meeting QoS requirements in terms of delay. Finally, we design distributed deep reinforcement learning approach to maximize the formulated reward function.  The simulation results show that our approach satisfies  more than 90\% vODUs resource constraints and network slice requirements.
	\end{abstract}

\section{Introduction}
\label{sec:introduction}
\subsection{Background and Motivations}
Flying cars were recently introduced in Urban Air Mobility (UAM) as an innovative concept for the transportation of people and goods \cite{ansari2021urban}. Flying cars are expected to become a reality in smart cities. Some essential projects for flying cars have recently been introduced, such as electric Vertical Take-Off and Landing (eVTOL) and Personal Aerial Vehicles (PAVs). The cruising altitude of the fying cars can reach around $300$ meters. The flying cars can fly at very high speeds, up to $300$ km/h. However, in terms of connectivity, using existing base stations in the cellular network without antennas adjustment is almost infeasible because antennas propagate towards the ground \cite{zaid2021technological}. As discussed in \cite{saeed2021wireless},  base stations can have additional antennas pointing toward the sky with omnidirectional coverage to address this challenge. Therefore, flying cars can operate within the coverage domains of ground base stations \cite{al2021connectivity}. In other words, the ground base stations can serve both ground-based cars and flying cars. Each car may need different services of different QoS and connectivity requirements such as high definition maps, remote pilot, autonomous driving, remote diagnosis, and infotainment contents.  Therefore, network slicing that enables virtualized networks on the same physical network can be an appropriate solution to fulfill the diverse requirements for services of flying and ground-based cars.   However, such heterogeneity of services per  each car cannot be effectively managed and efficiently mapped onto one slice. We need a slice per service. Also, some slices such as infotainment slice may serve flying and ground-based cars.

 Several prototypes have been designed for network slicing at the core network \cite{chahbar2020comprehensive}. However, Radio Access Network (RAN) slicing is still in the early stages. Therefore, this work focus on RAN slicing and consider the Open Radio Access Network (O-RAN) as a use case. However, O-RAN is not restrictive. O-RAN has been introduced to enable the intelligence and openness of RAN \cite{alliance2018ran}. O-RAN uses distributed intelligent controllers, where Near-Real-Time RAN Intelligent Controller (Near-RT RIC) enables training, testing, utilization, and updating machine learning. In contrast, Non-Real-Time RAN Intelligent Controller (Non-RT RIC) enables machine learning functionalities for policy-based guidance of applications and features. In O-RAN, there are three types of control loops. Loop $1$ operates at a time scale  less than $10$ msec. Loop $1$ can be employed for Resource Block (RB) scheduling in Transmission Time Interval (TTI). Loop $2$ operates at Near-RT RIC within the range of $10-1000$ msec. Loop $2$ can be appropriate for resource optimization. In Non-RT RIC, Loop $3$ operates at a time scale greater than $1000$ msec. Loop 3 can be employed for policies-based resource orchestration. Also, O-RAN supports O-RAN Central Unit Control Plane (O-CU-CP) and O-RAN Central Unit User Plane (O-CU-UP). O-RAN Central Units (O-CU-CP and O-CU-UP) interfaces with O-RAN Distributed Unit (O-DU) to provide services to edge devices via O-RAN Radio Units (O-RUs).  
\subsection{RAN slicing Challenges in Dealing with Car Services}
Considering slicing in RAN and O-RAN, the following are key challenging issues for serving the cars:
\begin{itemize}
	\item 
	Allocate radio resources and coordinate multiple RAN slices of multiple tenants who provide services to cars such that the required QoS is satisfied and Service Level Agreement (SLA)  is respected.
	\item Heterogeneity of services per each car such as ultra-low latency connectivity for autonomous driving/pilot, a high data rate for infotainment, and  an extremely high connection density for remote diagnosis.  One slice can not meet all required network features of the services needed by the car. 
	\item 
	High mobility of cars requires fast decisions in radio resources allocation. Therefore, a closed loop with real-time analytics is needed for taking appropriate and quick radio resource allocation decisions.
	\item  Satisfy slice requirements with high efficiency in finite radio resources. If radio resource sharing is too aggressive,  the slices can not meet the required QoS for car services, and this can cause services to degrade. 
\end{itemize}
\subsection{Contributions}
To address the aforementioned challenges, this work proposes two-level closed loops for managing RAN slice resources serving flying and ground-based cars. Our key contributions are summarized as follows:
\begin{itemize}
	\item 
	We propose an auction mechanism for allocating RBs to the tenants who provide services to flying and ground-based cars using slices. We assume the  RBs are limited, and tenants should compete to get them.
	\item 
	We propose one closed loop to create the slices associated to the services of tenants to vO-DUs for RBs scheduling purposes. We consider virtualized O-DU, where vO-DU is virtualized instance of O-DU.% and manages certain slices associated to the services to be provided to cars.
	\item 
	We propose another closed loop for intra-slices RB scheduling to serve flying and ground-based cars. Also, we design communication planning approach  that supports the proposed closed loop in RB scheduling.
	\item 
	We formulate a reward function that joins two closed loops and consider QoS fulfillment in terms of delay and workload changes. However, finding one solution that fits all two closed loops is a challenging issue. Therefore, we design distributed Reinforcement Learning (RL) approach that enables two closed loops to exchange experiences for maximizing the reward function. 
\end{itemize}
The rest of this paper is organized as follows. Section \ref{sec:LiteratureReview} discusses the related work, while  Section \ref{sec:system-model} presents the
system model. In Section \ref{sec:initialal_allocation}, we present initial resource allocation, while Section \ref{sec:ProblemFormulation} demonstrates the problem formulation. We  discuss the proposed solution in Section \ref{sec:ProposedSolution}. Section \ref{sec:PerformanceEvaluation} presents a performance evaluation. We conclude the paper in Section \ref*{sec:Conclusion}.
\begin{figure*}[t]
	\centering
	\includegraphics[width=1.7\columnwidth]{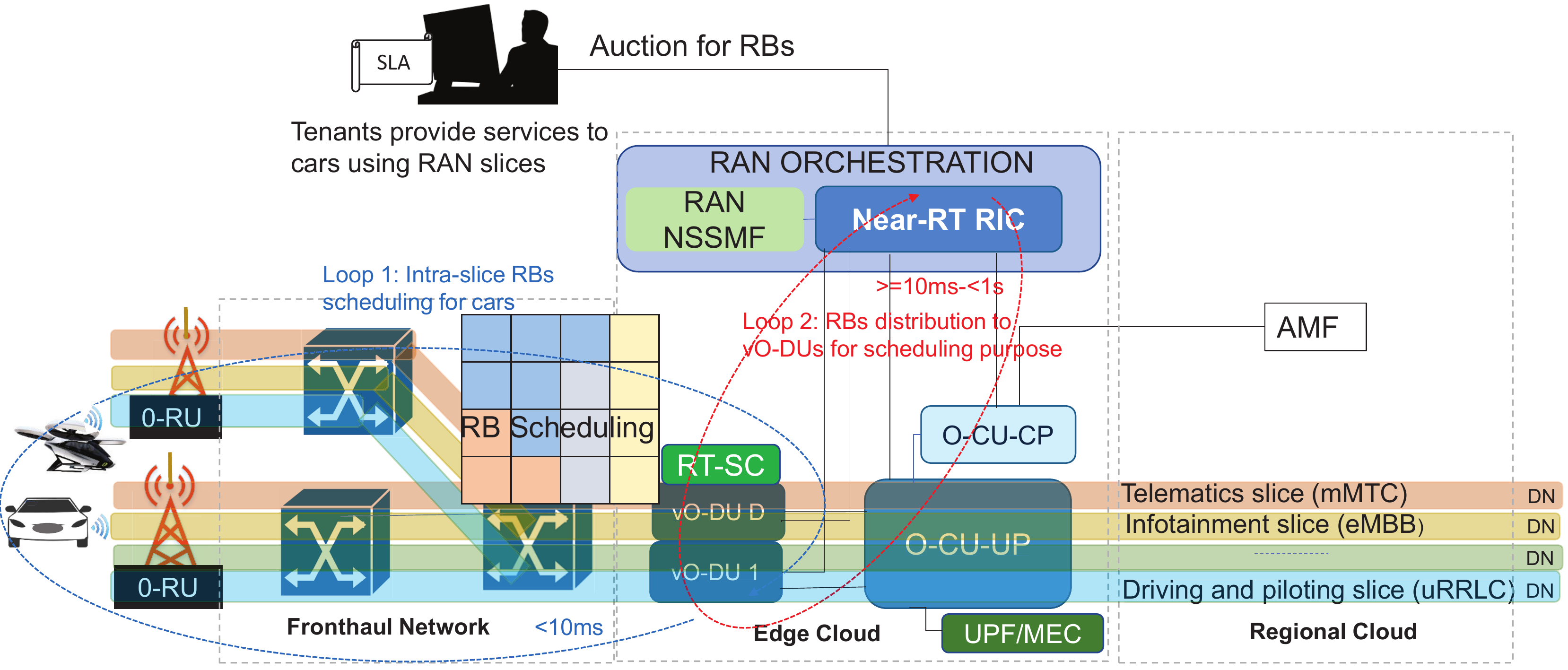}
	\caption{Illustration of our system model.}
	\label{fig:SystemModel}
\end{figure*}
\section{Literature Review}
\label{sec:LiteratureReview}
We group the existing related works into three categories: ($i$) network slicing in general, ($ii$) closed loops and RAN slicing, and ($iii$) RB allocation for RAN slices.

 \emph{Network slicing in general.}
 Network slicing has gained significant attention in literature \cite{chahbar2020comprehensive}. In this category, we discuss end-to-end network slicing.  The authors in \cite{salhab2019machine,fossati2020multi} proposed optimization framework to fine-grained resource allocation and Machine Learning (ML) approach to do traffic prediction. However, each use case scenario of 5G has its requirements in terms of energy, latency,  throughput, mobility, and reliability. Therefore, QoS requirements should be considered in network slicing. The authors in \cite{shu2020novel} proposed a QoS framework for network slicing that satisfies QoS of different 5G application scenarios. In \cite{kafle2018consideration}, the authors proposed ML approach for automation of network slice operations. In \cite{cheng2020safeguard}, the authors used deep learning and Lyapunov stability theories to enable the network to learn appropriate and safe slicing solutions. As decentralized deep learning solution,  the authors in \cite{liu2020edgeslice} proposed decentralized Deep RL (DRL) for edge computing networks that learns demands for network slices and orchestrates end-to-end resources.  In \cite{habibi2020should}, the authors discussed vertical industries with multiple use cases, where each use case is associated with diverging services and
 connectivity requirements. They used Vehicle to Everything (V2X)  communication slices as a slicing example. The authors in \cite{yuan2020airslice} proposed AerialSlice as a network slicing framework to handle unmanned aerial vehicle applications classified according to QoS requirements. In  \cite{garcia2019performance}, relying on the testbed,  the authors proposed a new 5G network slicing approach that provides connectivity to cars and trains using UAV. In view of the above discussed works, network slicing that considers distributed elements of O-RAN is new in the literature.
 
\emph{Closed loops and RAN slicing.} Here, we  discuss RAN slicing and application of closed loops in RAN slicing.
The authors in \cite{wu2020dynamic} discussed a new approach to satisfy the different QoS requirements for the Internet of vehicles services, where multiple slices are implemented at roadside units. In \cite{tamang2021architecting}, the authors proposed vehicle location-aware RAN slicing approach for mission-critical services. They used bandwidth reservation technique to serve vehicles using RAN slices. The authors in \cite{chang2018radio} proposed  RAN inter-slice resource partitioning and allocation as an optimization problem that facilitates inter-slice radio resource sharing.  The authors in  \cite{xie2019towards} discussed three closed loops to coordinate service management for network slices. Furthermore, in \cite{naik2022closed}, the author presented a closed loop deployment for automatic slicing assurance in 5G RAN to meet the SLA of each deployed slice. However, there are no mathematical modeling and solutions in these two works in  \cite{xie2019towards, naik2022closed}. The use and modeling of interconnected closed loops for network slicing is new in the literature.

\emph{RBs allocation for RAN slices.} The authors in \cite{kazmi2020distributed} proposed radio resource allocation using matching theory and auctions in a visualized wireless environment. The authors in \cite{abiko2020flexible} used DRL to perform RB allocation to the RAN slice, where each DRL agent manages one network slice. In \cite{azimi2021energy}, the authors presented an energy-efficient DRL-based solution for power and radio resources allocation in RAN slices. The author in \cite{albonda2019efficient} discussed off-line RL for allocating resources to RAN slices that serve enhanced mobile broadband (eMBB) and V2X services.

\emph{Novelties of this paper over related work.}
Our proposed approach  have several novelties over these prior approaches including: ($i$) while \cite{xie2019towards, naik2022closed} focused on one closed loop, we consider two closed loops that exchange experiences  for improving resource allocation; ($ii$) many related work focused on one type of vehicles in network slicing \cite{wu2020dynamic, tamang2021architecting, garcia2019performance, yuan2020airslice}, here, we combined flying and ground-based cars in network slicing;  ($iii$) managing RAN slices using two O-RAN closed loops in multi-tenants and multi-services environment of flying and ground-based cars is new and has not been tackled in literature.

\section{System model}
\label{sec:system-model}

\begin{table}[t]
	\caption{Summary of key notations.}
	\label{tab:table1}
	\begin{tabular}{ll}
		\toprule
		Notation & Definition\\
		\midrule
		$\mathcal{V}$ & Set of cars, $|\mathcal{V}|= V$\\
		$\mathcal{K}$ & Set of services, $|\mathcal{K}|= K$\\
		$\mathcal{M}$ & Set of O-RUs,  $|\mathcal{M}|= M$\\
		$\mathcal{B}$ & Set of RBs  $|\mathcal{B}|=B$\\
		$\mathcal{L}$ & Set of tenants  $|\mathcal{L}|=L$\\
		$\mathcal{C}$ & Set of slices $|\mathcal{C}|=C$\\
		$\mathcal{D}$ & Set of vO-DU $|\mathcal{D}|=D$\\
		$\lambda^v_{k,c}$ & Arrival rate of the packets for service $k$\\
		$J^{l, k}_b$ & Bid of tenant $l$ for RB\\
		$n^{l,k}_b$ & Number of RB needed by tenant $l$\\
		$\chi^m_{v}$& Distance between the car $v$ and O-RU $m$\\ 
		${I_v}$ & Speed of vehicle $v$\\
		$R^v_m$ & Achievable data rate of car $v$\\
		$r_t(\vect{y},\vect{z}, \vect{w})$ & Main reward function\\
     	$\Psi^d_{c,k}$ & Queue status parameter for service $k$\\
     	$\Omega^d_{c,k}$& Intra-slice orchestration parameter \\
		\bottomrule
	\end{tabular}
\end{table}

In our model depicted in Fig. \ref{fig:SystemModel}, we consider $\mathcal{V} = \{ 1, \dots,V	\}$  as a set of cars. In the cars, it includes both flying cars $\mathcal{V}_a$  and ground-based cars $\mathcal{V}_g$, such that $\mathcal{V} = \mathcal{V}_a \cup \mathcal{V}_g$. Each car $ v \in \mathcal{V}$ can require one or more services such as infotainment content, remote diagnosis, computation in  Multi-Access Edge Computing (MEC) server. We use $\mathcal{K} = \{ 1, \dots,K	\}$ as a set of services. Each service  $k \in \mathcal{K}$ needed by car $v \in \mathcal{V}$ is associated with  delay budget $\tau^v_k$, where delay budget is based on 5G QoS Identifier (5QI) defined in \cite{p1}. Each car requires network connection to get service. We assume  each car can be connected to O-RU  via a wireless network. We consider the Orthogonal Frequency Division Multiple Access (OFDMA) downlink scenario, where O-RU provides wireless connection to certain number of cars. We denote  $\mathcal{M} = \{ 1, \dots,M	\}$ as a set of O-RUs. In O-RUs includes
O-RUs of type RSU (Road-Side Unit), which support both
O-RU and V2X functionalities. 

The O-RUs and vO-DUs belong to Infrastructure Provider (InP), where InP has RBs $B$ at the cost of $\Gamma(B)$. We assume that the RBs are divisible for being allocated to the tenants who provide services to cars using the slices. We consider cars are subscribed to the slices of tenants. We denote   $\mathcal{L} = \{ 1, \dots,L	\}$ as a set of tenants. Each service of tenant can be mapped to specific slice types such as  enhanced Mobile Broadband (eMBB),  Ultra Reliable Low Latency Communications (URLLC), and  massive Machine Type Communications (mMTC). We use  $\mathcal{C} = \{ 1, \dots,C\}$ as a set of slices, where each slice manages one service. 
We use the auction to allocate RBs to the slices associated to the services of tenants. Near-RT RIC gets slice requirements from tenants via RAN Network Slice Subnet Management Function (NSSMF) and performs RBs allocation. In near real-time loop (loop 2 works in $10$ $ms$ to $1$ $s$), Near-RT RIC assigns RBs and slices to vO-DUs for management purpose. In the real-time loop (less than or equal to $10$ $ms$), each slice at vO-DU allocates RBs to cars. Here, we consider slicing at the core network and Data Network (DN) to be outside the scope of this paper. Also, we consider slice-aware  Access \& Mobility Management Function (AMF)  and O-CU-UP selection as future work.

\section{Initial Slice and Resource Block Allocation}
\label{sec:initialal_allocation}

\subsection{Resource Block Allocation to the Tenants}

 We consider RBs are limited. The tenants, who provide service to vehicles using slices, should compete to get RBs from InP. Therefore, InP makes RBs $B$ available to $L$ tenants of $K$ services for buying via auction. In the auction, we consider InP as a seller of  RBs and multiple tenants $L$ as buyers.

\begin{figure}[t]
	\centering
	\includegraphics[width=0.75\columnwidth]{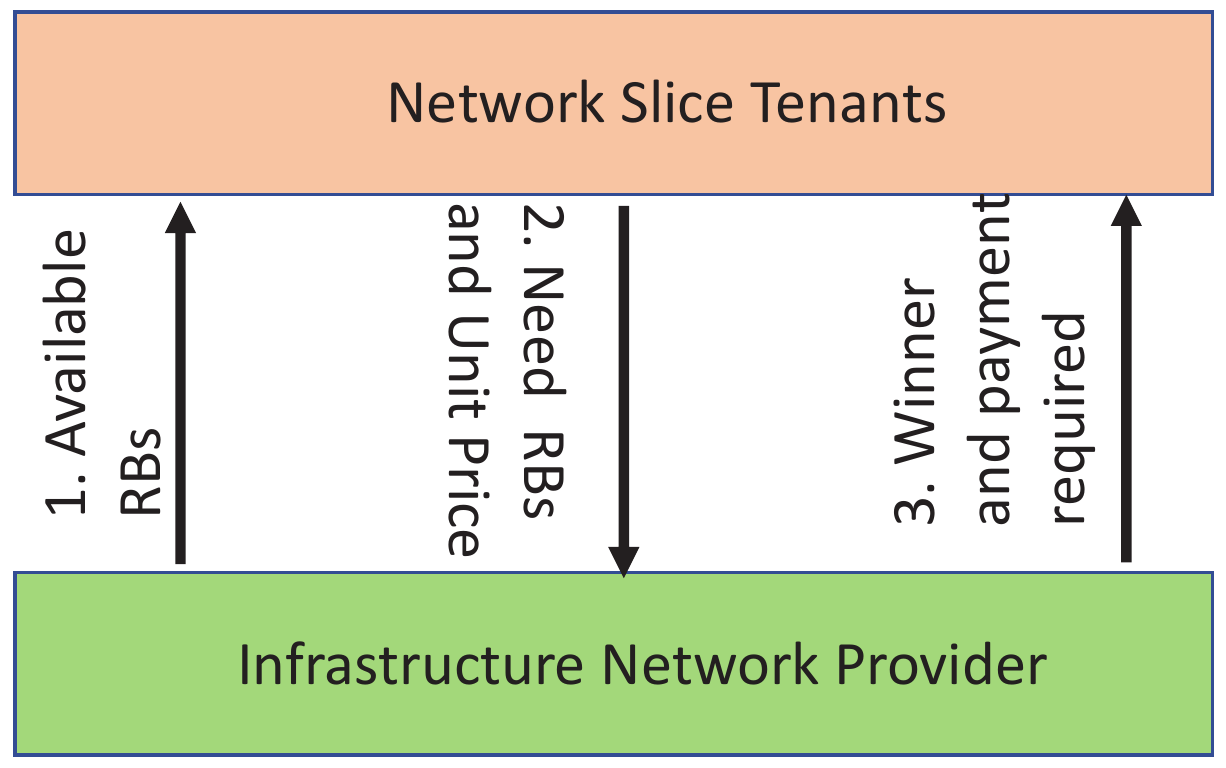}
	\caption{ Workflow of auction for RBs.}
	\label{AuctionModel}
\end{figure}

The workflow of Auction for RB (ARB) is presented in Fig. \ref{AuctionModel} and summarized as follows:
\begin{itemize}
	\item 
	\emph{Step 1:} The InP announces available RBs for auction to tenants $L$ and reserve price $b_p$ per unit of RB $b$. A reserve price $b_p$ represents minimum price that InP would accept from tenants per unit of RB $b$.
	\item 	
	\emph{Step 2:} In receiving available RBs for auction and  reserve price $b_p$, each tenant $l \in \mathcal{L} $ of service $k$ prepares and a submits bid $(J^{l, k}_b,n^{l,k}_b)$ to InP as demand for RBs. $J^{l, k}_b$ represents bid per unit of RB $b$ for service $k$  and
	$n^{l,k}_b$ represents initial number of RB $b$ needed for service $k$.
	\item 	
	\emph{Step 3:}
	InP collects all of the bids from the tenants and evaluates them. For $J^{l, k}_b \geq b_p$, the InP sorts the bids in descending order. Then, InP allocates the RBs to tenants starting with the tenant with highest bidding values. The InP calculates the payment $J^{l, k*}_b(n^{l,k}_b)$	that each winning tenant $l$ of service $k$ has to pay for RBs. Then, the InP declares the winning tenants and the winning price $J^{l, k*}_b(n^{l,k}_b)$.
\end{itemize}

% Each tenant  of service $k$ is asked to submit bid as demand for RB, then InP maps the vector of bids to RB and computes the payment that each tenant has to pay for getting requested RB. Furthermore, for helping the tenants to prepare their bids, InP announces the available  RB and unit price of RB. We use  $(J^{l, k}_b,n^{l,k}_b)$ to denote the bid tenant send to InP for RB needed for service $k$. $J^{l, k}_b$ is the bid per unit of RB and $n^{l,k}_b$ is the initial number of needed RB.
%Each service $k$ of a tenant receives the fraction of RB equals to:
% \begin{equation}
%	b^k=\frac{n^{l,k}_b}{\sum_{k \in \mathcal{K}}n^{l,k}_b},
%\end{equation}

%$
%\sum_{k \in \mathcal{K}}n^{l,k}_b \leq \sum_{d \mathcal{D}} b_d.
%$
% paper: Optimality of the Round-Robin Routing Policy

ARB helps the InP to choose  winning tenants that submitted bidding values that maximize its revenue and the social welfare. In ARB, we consider that each tenant $ l \in \mathcal{L}$ submits its bid for RB $b\in \mathcal{B}$ without knowing the bidding values of other tenants. Also, each tenants $l \in \mathcal{L}$ can  submit one bid per service. We consider that each tenant $l$ has its own valuation for RB $b$ denoted $\Upsilon_{l,k}(n^{l,k}_b)$. Here, $\Upsilon_{l,k}(n^{l,k}_b)$ is given by:
\begin{equation}
	\setlength{\jot}{10pt}
	\Upsilon_{l,k}(n^{l,k}_b) =
	\begin{cases}
		\iota^{l,k}_b
		n^{l,k}_b,\;\text{if the tenant $l$ paricipates  in ARB,} \\
		0, \; \text{otherwise,}
	\end{cases}
\end{equation}
where $\iota^{l,k}_b$ is the true valuation of tenant $l$  for service $k$ that requires RB $b$.
However, when tenant $l$ does not participate in the ARB, its true valuation is $0$. On the other hand, the valuation  $\Gamma(B)$ of the InP is defined using reserved price $b_p$ such that 
$\Gamma(B)=B b_p$. InP sets $b_p$ that ensures its revenue does not become negative. In other words, its revenue covers its CAPEX and OPEX associated to RBs. 

In our action, we choose Vickrey Clarke Groves (VCG) mechanism \cite{caminati2016vickrey} over other auction mechanisms because  VCG mechanism enables welfare maximization of all tenants and guarantees a truthful outcome. VCG enables to achieve better efficiency in RBs allocation and competition between tenants. It allows optimal price  $J^{l, k*}_b(n^{l,k}_b)$ for RB to come from the competition. To apply the VCG in our auction, we define the maximum valuation $\Upsilon_L(n^{l,k}_b)$ of all tenants with bidding values $J^{l, k}_b \geq b^k_p$ as follows:
\begin{equation}
	\begin{aligned}
		\Upsilon_L(n^{l,k}_b)=	\underset{J^{l, k}_b \geq b_p}{\text{argmax}}
		\sum_{l \in \mathcal{L}}^{}J^{l, k}_b n^{l,k}_b.\\
	\end{aligned}
	\label{eq:winner_determination1}
\end{equation} 
In the VCG, each tenant $l$ should pay for the damage it may cause on other tenants by participating in the ARB. Therefore, we compute the total valuation $\Upsilon_{-l}(n^{j,k}_b)$  without each
tenant $l$, where $\Upsilon_{-l}(n^{j,k}_b)$ is given by:
\begin{equation}
	\begin{aligned}
		\Upsilon_{-l}(n^{j,k}_b)=	\underset{J^{j, k}_b \geq b_p}{\text{argmax}}
		\sum_{j \in \mathcal{L}\setminus \{l\}^{}}J^{j, k}_b n^{j,k}_b.\\
	\end{aligned}
	\label{eq:winner_determination2}
\end{equation}
From (\ref{eq:winner_determination1}) and (\ref{eq:winner_determination2}), we can compute the price $J^{l, k*}_b(n^{l,k}_b)$ that each tenant $l$ of service $k$ has pay to InP  as follows: 
\begin{equation}
	\begin{aligned}
		J^{l, k*}_b(n^{l,k}_b)= \Upsilon_{-l}(n^{j,k}_b) -\sum_{j\neq l}^{}J^{j, k}_b n^{j,k}_b.\\
	\end{aligned}
	\label{eq:winner_determination_price}
\end{equation} 

\begin{definition}[Tenant Utility]
	In ARB, in which tenant submit a bid $(J^{l, k}_b,n^{l,k}_b)$, if the tenant $l$ wins the ARB,
	it pays $J^{l, k*}_b(n^{l,k}_b)$ to InP. Otherwise, if tenant $l$ loses the ARB, it  pays nothing. Therefore, the utility $U_{l,k}$ of any tenant $l$ of service $k$ is given by:
\end{definition}
\begin{equation}
	\label{eq:utility0}
	\setlength{\jot}{10pt}
	U_{l,k} =
	\begin{cases}
		J^{l, k*}_b(n^{l,k}_b) - \Upsilon_{l,k}(n^{l,k}_b) ,\; \text{if tenant l $\in \mathcal{W}$ win ARB}\\
		0, \text{ otherwise,}
	\end{cases}
\end{equation}
where $\mathcal{W}$ is the set of the winners. We consider each tenant  will  participate in ARB if and only if $	J^{l, k*}_b(n^{l,k}_b) \geq \Upsilon_{l,k}(n^{l,k}_b)$. In other words, a tenant will participate in ARB when its utility is not negative.

\begin{definition}[Individual Rationality]
	\label{individualRationality}
	ARB is individually rational if and only if no tenant $l \in \mathcal{L}$ receives negative
	utility, i.e.,  $U_{l,k}$ is not negative ($ U_{l,k} \geq 0$).
\end{definition}
\begin{definition}[Truthfulness] 
	\label{Truthfulness}
	ARB is truthful if and only if, for each tenant $l \in \mathcal{L}$, bidding the truth value $\iota^{l,k}_b=J^{l, k}_b$ is the dominant strategy. In other words, bidding  $\iota^{l,k}_b$ that maximizes the utility of each tenant $l \in \mathcal{L}$ given for all possible bidding values is the dominant strategy.			
\end{definition}
\begin{theorem}The ARB is truthful.
	\label{RA-truthful} 
\end{theorem}
\begin{proof} 
We consider that each tenants $l \in \mathcal{L}$ wins the ARB by submitting its true valuation, i.e.,  $\iota^{l,k}_b=J^{l, k}_b$. Also, ARB satisfies monotonicity and critical payment conditions of  truthful bidding defined in \cite{blumrosen2007combinatorial}.
\begin{itemize}
	\item 
	\emph{Monotonicity:} Let us consider a scenario of two tenants $l$ and $l'$ submitted bidding values $J^{l, k}_b$ and $J^{l', k}_b$ for service $k \in \mathcal{K}$, where  $J^{l, k}_b > J^{l', k}_b$. ARB chooses bidding value that maximizes total valuation in descending order of the bidding values. Therefore, $J^{l, k}_b$ will give more chance tenant $l \in \mathcal{L}$ to win ARB over $J^{l', k}_b$ because $J^{l, k}_b > J^{l', k}_b$. 
	\item 
	\emph{Critical payment:} In ARB, the payment of winner is based on its bidding value and the bidding values  of other tenants, where VCG tries to maximize social welfare. The ARB  makes tenants $l \in \mathcal{L}$ with maximum bidding value $J^{l, k}_b$ as the winner whatever other bidding values such as $J^{l', k}_b$, and winner $l \in \mathcal{L}$ pays $	J^{l, k*}_b(n^{l,k}_b) \leq J^{l, k}_bn^{l,k}_b$.
\end{itemize}
\end{proof} 
\begin{theorem}The ARB is individually rational.
	\label{RA-rational} 
\end{theorem}
\begin{proof}  Considering Definition \ref{individualRationality} and individually rational condition defined in \cite{blumrosen2007combinatorial}, ARB becomes individually rational when no tenant receives negative utility. Based on the above Theorem \ref{RA-truthful} and (\ref{eq:utility0}), ARB  makes tenant $l \in \mathcal{L}$ with maximum bidding value $J^{l, k}_b$ as the winner whatever other bidding values and pays $J^{l, k*}_b (n^{l,k}_b) \leq J^{l, k}_bn^{l,k}_b$. Otherwise, based (\ref{eq:utility0}), tenant who does not win ARB receives zero utility ($U_{l,k} = 0$). Therefore, $ U_{l,k}  \geq 0$.
\end{proof}

The above ARB can be designed as Total Revenue Maximization (TRM)  problem, where TRM is expressed as follows: 
\begin{subequations}
	\label{eq:sub11}
	\begin{align}
		&\underset{\vect{x}}{\text{maximize}}\ \  \sum_{k\in \mathcal{K}}\sum_{l\in \mathcal{L}} x^{l, k}_b n^{l,k}_bJ^{l, k}_b
		\tag{\ref{eq:sub11}}\\
		&\text{subject to:}\nonumber\\
		& \sum_{k\in \mathcal{K}}\sum_{l \in \mathcal{L}}^{}x^{l, k}_b n^{l,k}_b\leq B, \;  \forall b \in \mathcal{B}\label{first:aa},\\
		&  x^{l, k}_bJ^{l, k}_b \geq b_p,\label{first:bb}\\
		& x^{l, k}_b\in \{0,1\} \label{first:dd}.
	\end{align}
\end{subequations}
In TRM problem (\ref{eq:sub11}), the RBs needed to be allocated to tenants must be less than the total RBs. In (\ref{first:bb}), the bidding value of the tenant should be greater or equal to the reserve price of InP. In (\ref{first:dd}), we use $x^{l, k}_b$ as binary decision variable, where $x^{l, k}_b=1$ if tenant $l$ submit bid $J^{l, k}_b$ and wins the auction, and $x^{l, k}_b=0$ otherwise.

TRM problem is an Integer Linear Programming (ILP) problem. To handle (\ref{eq:sub11}),  we propose  an algorithm (Algorithm \ref{algo:WD}) for Winner and Price Determination. Algorithm \ref{algo:WD}  is based on the VCG mechanism. 
The inputs of Algorithm \ref{algo:WD} include a set of tenants $\mathcal{L}$, set of services $\mathcal{K}$, available RBs $B$ for auction,  vector of bids $\vect{J}_b$, vector of the number of RBs needed $\vect{n}$. At the line $3$, the algorithm initializes the parameters of the auctions including set of winners $\mathcal{W}$ and set of tenants $\mathcal{W}'$ who do not win the auction. Then, the algorithm performs iterations for winner and price determination until all RBs $B$ are allocated to the tenants or no more tenants need RBs. The outputs of the Algorithm \ref{algo:WD} are set of winning tenants  $\mathcal{W}$, vector $\vect{x}$ of winning decision variables, and vector of  $\vect{J}^*$ payments. We assume that $J^{l, k*}_b(n^{l,k}_b)$ is the flat price that the tenant $l$ and InP agreed for RBs of slice associated to service $k$ during the auction. Once the tenant  RB usage passes the initial number of RB $n^{l,k}_b$ requested in the auction, i.e., cap, InP does not stop the tenant service, but InP introduces a flat rate increase described in \cite{chiang2012networked}. However, we consider a flat rate increase to be outside the scope of this paper. Also, the auction is performed outside the closed loops. In other words, the auction helps to get RBs that will be managed using closed loops.
\begin{algorithm}
	\caption{: Winner and price determination for ARB.}
	\label{algo:WD}
	\begin{algorithmic}[1]
		\STATE {\textbf{Input:} $\mathcal{K}$, $\mathcal{L}$, $B$, $b_{p}$,  $\vect{J}_b$, $\vect{n}$;}
		\STATE {\textbf{Output:} $\mathcal{W},\vect{x}, \vect{J}^*$\\ //Initialization; }	
		\STATE {$\mathcal{W}\gets  \emptyset$, $\vect{J}^*\gets\emptyset$, $\mathcal{W}'\gets  \emptyset$, $\vect{x} \gets  (0,\dots,0)$, $\Upsilon_L(n^{l,k}_b) \gets 0$,  $\Upsilon_{-l}(n^{j,k}_b) \gets 0$, $\Upsilon(b) \gets 0$; } 
		\WHILE{ $\vect{n}\neq  \emptyset$ and $J^{l, k}_b \geq b_p>0,$}
		\STATE {  $\vect{J}_{b}$ $\gets J^{l, k}_b $  ;} 
		\STATE { Sort $\vect{J}_{b}$ in decreasing order;} 
		\REPEAT
		\STATE{Find a tenant $l$ that has the maximum bid $J^{l, k}_b$ ($max(\vect{J}_{b})$) as a winner;}
		\STATE{$\vect{J}^*\gets J^{l, k}_b;$}
		\STATE{$\Upsilon_L(n^{l,k}_b)= \Upsilon_L(n^{l,k}_b) +J^{l, k}_b n^{l,k}_b$;}	
		\STATE{$\mathcal{W}\gets \mathcal{W} \cup \{l\}$;}	
		\STATE{$\mathcal{L}_0\gets \mathcal{L} \setminus \{l\}$;}	
		\STATE{$x^{l, k}_b\gets 1$;}	
		\STATE {$\vect{x}\gets x^{l, k}_b;$}
		\STATE{$B=B-n^{l,k}_b$;}		
		\UNTIL{$B=0$ or $\mathcal{L}=\emptyset$;}
		\ENDWHILE
		\STATE{ Reset $B$ and $\mathcal{L}$;}	
		\REPEAT
		\STATE{Find a tenant $j \in  \mathcal{L}' = \mathcal{L}_0 \cup \mathcal{W} \setminus  \{l\}$ that has the
			maximum bid $J^{j, k}_b$ ($max(\vect{J}_{b})$) when each tenant $l \in  \mathcal{W}$ is
			not participating in the auction;}
		\STATE{$\Upsilon_{-l}(n^{j,k}_b) \gets \Upsilon_{-l}(n^{j,k}_b) +J^{j, k}_b n^{j,k}_b$;}	
		\STATE{$\mathcal{W}'\gets \mathcal{W}' \cup \{j\}$;}	
		\STATE{$\mathcal{L}'\gets \mathcal{L}' \setminus \{j\}$;}	
		\STATE{$x^{j, k}_b\gets 0$;}
		\STATE {$\vect{x}\gets x^{j, k}_b;$}
		\STATE{$B=B-n^{j,k}_b$;}		
		\UNTIL{$B=0$ or $\mathcal{L}'=\emptyset$;}
		\WHILE{ $j\neq l \in \mathcal{L}_1 = \mathcal{W} \cup \mathcal{W}'$}
		\STATE{Find a tenant $j$ that has the
			maximum bid $max(\vect{J}_{b})$ when tenant $j \in  \mathcal{W'}$ and $l \in  \mathcal{W};$}
		\STATE{$\Upsilon(b) \gets \Upsilon(b) +J^{j, k}_b n^{j,k}_b$;}
		\ENDWHILE
		\STATE{	$^{l, k*}_b(n^{l,k}_b)= \Upsilon_{-l}(n^{j,k}_b) -\Upsilon(b);$}
		\STATE {$\vect{J}^*\gets J^{l, k*}_b(n^{l,k}_b);$}
		\STATE {\textbf{Return:} $\mathcal{W},\vect{x}, \vect{J}^*$.}		
	\end{algorithmic}
\end{algorithm}

\begin{theorem}Computational complexity of ARB is $O(n^2)$
	\label{RA-TPM} 
\end{theorem}
\begin{proof} 
	In the Algorithm \ref{algo:WD}, we have while loop at lines $4-17$ that  performs $n$ iterations for checking submitted bids ($J^{l, k}_b \geq b_p>0$), where $n$ is the size of the vector $\vect{J}_b$. Inside the while loop, we have another loop at lines $7-16$ for allocating RBs to the tenants starting from the tenant with maximum bidding value and this loop takes $n$ iterations. We have third loop at lines ($19-27$) for finding the winners if each tenant with maximum bidding value does not participate in ARB,  which takes $n-1$ iterations. The last loop is at lines $(28-31$) for calculating total evaluation and it takes $n$ iterations. As result, the Algorithm \ref{algo:WD} takes $n^2 + n- 1 + n$ iterations. In conclusion, the
	computational complexity of RA is $O(n^2)$, which is linear time.
\end{proof}

\subsection{RBs Distribution to vO-DUs for  Scheduling Purpose}
In closed loop two, initially, InP assigns RBs $B$ to vO-DUs equally such that $B=\sum_{d=1}^{D}b_d$, where   $b_d=\lfloor\frac{B}{D}\rfloor $ is the RB assigned to each vO-DU $d$. After the auction, InP creates slices $C$ associated to $K$ services at vO-DUs and assigns  RBs to slices. InP uses  round-robin policy \cite{elamaran2019greedy} to create each slice $c \in \mathcal{C}$ associated to service $k \in \mathcal{K}$ of each winning tenant $l$ at vO-DU. The round-robin policy  cyclically create slices associated with services to vO-DUs starting from vO-DU $1$  such that $\sum_{k=1}^{K_d} b^{c,k}_d \leq b_d$, where $b^{c,k}_d$ is RBs of  each slice $c$ at each vO-DU $d$ for service $k$. $K_d$ represents the number of services at vO-DU $d$ and  $ b^{c,k}_d =y^{c,d}_{b,k} x^{l, k}_b	n^{l,k}_b$. Furthermore, we define $y^{c,d}_{b,k}$ as decision variable indicating whether slice $c$ of service $k$ has assigned radio resource at vO-DU $d$, where $y^{c,d}_{b,k}$ is given by:
\begin{equation}
	\label{eq:mappingslice_v0du}
	\setlength{\jot}{10pt}
	y^{c,d}_{b,k}=
	\begin{cases}
		1,\; \text{if slice $c$ of service $k$ has assigned RBs $\,$}\\ \;\;\;\;\text{at vO-DU $d$,}\\
		0, \;\text{otherwise.}
	\end{cases}
\end{equation}
To ensure that each slice $c$ of service $k$ is created at one vO-DU, InP imposes the following constraint:
\begin{equation}
	\begin{aligned}
		\sum_{c\in \mathcal{C}}y^{c,d}_{b,k} \leq 1 , \;  \forall d, b, k.	\end{aligned}
\end{equation}

\subsection{Intra-slices RBs Scheduling for Cars}
In closed loop 1, we consider vO-DUs are connected to O-RUs via wired fronthaul network, where O-RUs serve $V$ cars available in their coverage areas.
Based on chosen numerology $i$, each RB $b^{c,k}_d$  is partitioned
into $f^{c,k}_{i,d}$ number of sub-bands, indexed by $\mathcal{F}^{c,k}_{i,d} = \{1, 2, \dots ,F^{c,k}_{i,d}\}$ in the frequency-domain and $t^{c,k}_{i,d}$ number of TTIs, indexed by $\mathcal{T}^{c,k}_{i,d} = \{1, 2, \dots , T^{c,k}_{i,d}\}$ in the time-domain. Therefore, a total $F^{c,k}_{i,d} \times T^{c,k}_{i,d}$ number of RBs are available for the service $k$ using numerology $i$. RBs scheduling  can be modeled using  perfect Channel State Information (CSI). However, in practice, it is challenging to obtain perfect CSI due to some limitations such as delayed feedback.  As described in \cite{korrai2020joint}, the channel coefficient between the O-RU and scheduled cars on the RB
$(t^{c,k}_{i,d}, f^{c,k}_{i,d})$ of numerology $i$ is modeled as:
\begin{equation}
	h^{v,k}_{t_i,f_i}= \tilde{h}^{v,k}_{t_i,f_i}+ e^{v,k}_{t_i,f_i},
\end{equation}
where $\tilde{h}^{v,k}_{t_i,f_i}$ and $e^{v,k}_{t_i,f_i}$ represent the estimated CSI and estimated error, respectively. Using $h^{v,k}_{t_i,f_i}$, the achievable  achievable SNR at the cars $v$ on the RB $(t^{c,k}_{i,d}, f^{c,k}_{i,d})$ becomes:
\begin{equation}
	\label{eq:decisoF_nariable}
	\delta^v_{t_i,f_i} = 
	\frac{y^{c,d}_{b,k}|h^{v,k}_{t_i,f_i}|^2 \tilde{p}_{t_i,f_i} \chi^m_{v}}{\sigma_v^2},\; 
\end{equation}
where 
$ \tilde{p}_{t_i,f_i}$ is the allocated power to the each RB $(t^{c,k}_{i,d}, f^{c,k}_{i,d})$, $\chi^m_{v}$
is the distance between the car $v$ and O-RU $m$ and $\sigma_v^2$
is the noise power. 
\begin{figure}[t]
	\centering
	\includegraphics[width=1.0\columnwidth]{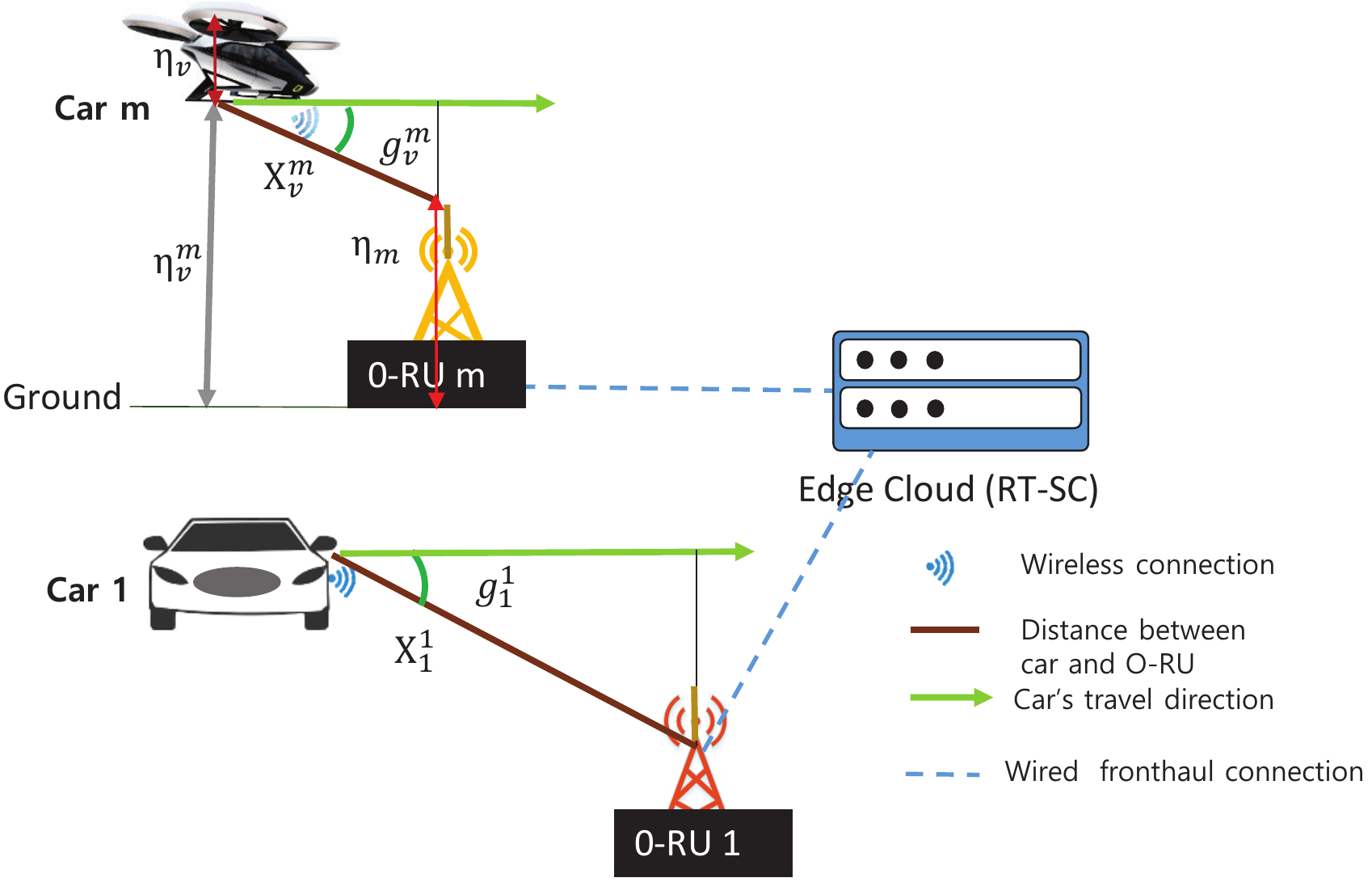}
	\caption{Communication planning for the cars.}
	\label{ComPlan}
\end{figure}

As shown in Fig. \ref{ComPlan}, due to car mobility, the distance   $\chi^m_{v}$ keeps changing. Therefore, the combination of global navigation satellite systems (GNSS) such as {GPS} and {GLONASS} can be applied to find $\chi^m_{v}$. The same approach was applied in \cite{sekaran2020modeling, ndikumana2020deep, ndikumana2019self}.  Furthermore,  we consider the distance of a flying car from the earth and height the O-RU, where O-RU  has antennas pointing toward the sky for aerial coverage to serve flying cars.  As  described in \cite{saeed2021wireless}, $\chi^m_{v}$  for the flying cars can be calculated as follows:
\begin{equation}
	\label{eq:distanceAerialvehicle_O-RU}
	\begin{aligned}
		\chi^m_{v}=\sqrt{\eta_m^v + (\eta_v-\eta_m)^2},  \;\forall v \in \mathcal{V}_a,
	\end{aligned}
\end{equation}
where $\eta_m$ is the height of O-RU $m$, $\eta_m^v$ is the estimated flying car to  O-RU $m$ projection distance on the ground, and $\eta_v$ is the estimated height of the flying car.

We consider the list of O-RUs is a priori known at edge cloud, i.e., at Real-time Slice Controller (RT-SC).
 RT-SC can calculate the remaining distance  $\varsigma^v_m$ of each car $v$ to reach area $\Lambda_m$ covered by each nearby O-RU $m$, where $\varsigma^v_m$ is given by:
\begin{equation}
	\label{eq:distancevehicle_O-RU}
	\begin{aligned}
		\varsigma^v_m= \chi^m_{v} cos g^m_v.
	\end{aligned}
\end{equation}
We use $g^m_v$ as an estimated angle between the trajectory of movement of  car $v$ and the  line from  O-RU $m$.
By using $\varsigma^v_m$, the RT-SC can compute the probability  $p^m_v$ that O-RU $m$ can serve car $v$ using wireless communication such that:
\begin{equation}
	\setlength{\jot}{10pt}
	p^m_v=
	\begin{cases}
		1,\; \text{if $\varsigma^v_m=0$ and $\tau^m_{v} \leq \tau^v_k$}, \\
		0,\;\text{otherwise.}
		\label{eq:probability_O-RU}
	\end{cases}
\end{equation}
When $\varsigma^v_m=0$, the  car $v \in \mathcal{V}$ reaches the area $\Lambda_m$ covered by  O-RU $m$.
We define $\tau^m_{v}$ as the time required by  car $v$ to leave the coverage area of O-RU $m$, where $\tau^m_{v}$ is given by:
\begin{equation}
	\setlength{\jot}{10pt}
	\tau^m_{v}=\frac{ \Lambda_m}{I_v},
\end{equation}
where ${I_v}$ is the estimated speed of car $v$.  When $\tau^m_{v} \leq \tau^v_k$, the car can easily use O-RU $m$ for wireless communication and meet delay budget $\tau^v_k$. Otherwise, when $\tau^m_{v}> \tau^v_k$, our approach can select the next O-RU to use that can satisfy the delay budget. However, we consider O-RU handover for flying and ground-based cars as future work.

According to
Shannon’s theory, the achievable data rate for the car $v$ on the RB $(t^{c,k}_{i,d}, f^{c,k}_{i,d})$ can be written as:
\begin{equation}
	\label{eq:data_rate}
	\begin{aligned}
		R^{v,m}_{t_i,f_i} =\omega^m_{t_i,f_i}p^m_v\log_2\left(1 + \delta^v_{t_i,f_i}\right),  \;\forall v \in \mathcal{V},
	\end{aligned}
\end{equation}
where $\omega^m_{t_i,f_i}$ is the bandwidth of the RB with numerology
$i$. Then, the data rate of each car $v$ can be computed as:
\begin{equation}
	\label{eq:data_rate_all}
	\begin{aligned}
		R^v_m =\sum_{i=1, \dots, 4}\sum_{t^{c,k}_{i,d}=1}^{T^{c,k}_{i,d}} \sum_{f^{c,k}_{i,d}=1}^{F^{c,k}_{i,d}} z^{v,m}_{t_i,f_i} R^v_{t_i,f_i}, 
	\end{aligned}
\end{equation}
where $z^{v,m}_{t_i,f_i}$ is binary decision variable indicates whether car $v$  uses  RB $(t^{c,k}_{i,d}, f^{c,k}_{i,d})$ of numerology $i$ at O-RU $m$, where $z^{v,m}_{t_i,f_i}$  is given by:
\begin{equation}
	\label{eq:RB_allocation_variable}
	\setlength{\jot}{10pt}
	z^{v,m}_{t_i,f_i}=
	\begin{cases}
		1,\; \text{If $p^m_v=1$ and RB $(t^{c,k}_{i,d}, f^{c,k}_{i,d})$ is allocated}\\
		\;\;\;	\; \text{ to car $v$,}\\
		0, \;\text{otherwise.}
	\end{cases}
\end{equation}
To comply with the requirement of OFDMA system, where each RB $(t^{c,k}_{i,d}, f^{c,k}_{i,d})$ can only be allocated to a single car, we impose the following orthogonality constraint:
\begin{equation}
	\begin{aligned}
		\sum_{u\in \mathcal{V}}z^{v,m}_{t_i,f_i} \leq 1 , \;  \forall v, t^{c,k}_{i,d}, f^{c,k}_{i,d}.	\end{aligned}
\end{equation}

\section{Problem Formulation for Two-level closed loops} 
\label{sec:ProblemFormulation}

The previous section discussed the two closed loops in initial RBs distribution and scheduling. This section discusses RBs distribution and scheduling feedback.

\emph{Feedback for closed loop 1:} After RBs scheduling for cars, we monitor RBs utilization. We consider $\lambda^v_{k,c}$ as the arrival rate of the packets for each service $k$ needed by car $v$. RT-SC maps incoming packets with vO-DU that manages  slice $c$ of service $k$. Each service has its queue, where queuing delay can be modeled with M/M/1 queuing system, where queuing delay $q_c^{v,k}$ can be expressed as follows:
\begin{equation}
	q_c^{v,k}= \frac{z^{v,m}_{t_i,f_i}}{\lambda^v_{k,c}- \mu^v_{k,c}},
\end{equation}
where $\mu^v_{k,c}$ represents the service rate. $w^{v,d}_{k,c} $ is binary decision variable indicating whether or not packet is  assigned to slice $c$ associated to service $k$ at vO-DU $d$, where $w^{v,d}_{k,c} $ is given by:
\begin{equation}
	\label{eq:map_allocation_variable}
	\setlength{\jot}{10pt}
	w^{v,d}_{k,c} =
	\begin{cases}
		1,\; \text{if packet is  assigned to slice $c$ associated to}\\
		\; \; \;\;   \; \text{service $k$ at vO-DU $d$,}\\
		0, \;\text{otherwise.}
	\end{cases}
\end{equation}
Furthermore, we consider buffer $\tilde{\beta}^d_{c,k}$ associated to service $k$ that uses slice $c$ at vO-DU $d$. Then, we introduced queue status parameter $\Psi^d_{c,k}$ associated to each service $k$ and buffer threshold $\beta^d_{c,k}$, where $\Psi^d_{c,k}$ can  dynamically computed  as follows:
\begin{equation}
	\setlength{\jot}{10pt}
	\Psi^d_{c,k} =\max \{(\tilde{\beta}^d_{c,k}-{E[\lambda^v_{k,c}]}), \beta^d_{c,k} \},
\end{equation}	
where $E[\lambda^v_{k,c}]$ is the expected number of packets in queue or queue occupancy for service $k$.

Besides queuing delay and status, we consider transmission and prorogation delays. We assume that each packet of the car $v$ passes through fronthaul and wireless network. Let us consider  $o_c^{v,k}$ as the size of the packet. The transmission delay for the wireless network between car and O-RU becomes:
\begin{equation}
	\tau_{c,k}^{v\rightarrow m}= \frac{o_c^{v,k}}{R^v_m}.
\end{equation}
Furthermore, the transmission delay $\tau_{c,k}^{m_v\rightarrow d}$ for fronthaul between O-RU $m$ and  vO-DU $d$ can be expressed as follows:
\begin{equation}
	\tau_{c,k}^{m_v\rightarrow d}=\frac{o_c^{v,k}}{\varpi_{m,d}},
\end{equation}
where $\varpi_{m,d}$ is the capacity of fronthaul link between O-RU $m$ and vO-DU $d$.  
The propagation delay $\tau^{m\rightarrow d}$ can be expressed as follows:
\begin{equation}
	\tau^{m\rightarrow d}=\frac{\rho^{m\rightarrow d}}{\kappa},
\end{equation}
where $\rho^{m\rightarrow d}$ is the length of fronthaul link $(m,d)$ and $\kappa$ is the propagation speed.     
The end-to-end delay can be expressed as follows:
\begin{equation}
	\tau_{c,k}^{v}= q_c^{v,k}+ \tau_{c,k}^{v\rightarrow m} +\tau_{c,k}^{m_v\rightarrow d} + \tau^{m\rightarrow d}.
\end{equation}

We consider $\tau_{c,k}^{v}$ as feedback for the loop $1$, where $\tau_{c,k}^{v}$ should satisfy delay budget constraint  $\tau_{c,k}^{v} \leq \tau^v_k$.

To evaluate intra-slices RB allocation using closed loop $1$, we defined network slice requirement satisfaction $\varphi^c_k$.  $\varphi^c_k$ measures whether or not each slice $c$ of  service $k$ satisfies delay budget $\tau^v_k$. The $\varphi^c_k$ is expressed as:
\begin{equation}
		\label{eq:network_slice_requirement_satisfaction}
	\varphi^c_k=\frac{\sum_{v=1}^{V_k}z^{v,m}_{t_i,f_i} \xi^v_{c,k}}{V_k},
\end{equation}
where  $\mathcal{V}_k$ is a set of cars that use service $k$ and  $\xi^v_{c,k}$ is the delay budget fulfillment parameter. $\xi^v_{c,k}$ is given by:
\begin{equation}
	\label{eq:satisfaction}
	\setlength{\jot}{10pt}
	\xi^v_{c,k}=
	\begin{cases}
		1,\; \text{if $\tau_{c,k}^{v} \leq \tau^v_k$}\\
		0, \;\text{otherwise.}
	\end{cases}
\end{equation}
To update initial RBs allocation for cars, we define intra-slice orchestration parameter $\Omega^d_{c,k}$ for close loop $1$, where $\Omega^d_{c,k}$ is given by:
\begin{equation}
	\setlength{\jot}{10pt}
	\Omega^d_{c,k}=
	\begin{cases}
		\frac{\tilde{\beta}^d_{c,k}}{\beta^d_{c,k}},\; \text{{if  $\Psi^d_{c,k} = \beta^d_{c,k}$},}\\
		\frac{\beta^d_{c,k}}{\tilde{\beta}^d_{c,k}},\; \text{if  $\Psi^d_{c,k} > \beta^d_{c,k}$,}\\
		0,\; \text{ if  $\Psi^d_{c,k} = \tilde{\beta}^d_{c,k}$,}\\
		1,\; \text{otherwise}.
	\end{cases}
\label{eq:probability_RSU}
\end{equation}
For close loop $1$, when  $\Psi^d_{c,k} = \beta^d_{c,k}$, we consider that there are many incoming packets for slice $c$ associated to service $k$. In this scenario vO-DU $d$ needs  performs  slice resource scale-up with $\Omega^d_{c,k}=\frac{\tilde{\beta}^d_{c,k}}{\beta^d_{c,k}}$ rate. Also, if  $\Psi^d_{c,k} > \beta^d_{c,k}$, the vO-DU $d$  needs to  perform  slice resource scale-down  with $\Omega^d_{c,k}=\frac{\beta^d_{c,k}}{\tilde{\beta}^d_{c,k}}$ rate  because the RB are under utilized ($E[\lambda^v_{k,c}]$ is small). When   $\Psi^d_{c,k} = \tilde{\beta}^d_{c,k}$, there is no demands for slice $c$ associated to service $k$, vO-DU $d$ can terminate RB allocation to that slice using $\Omega^d_{c,k}=0$ because $E[\lambda^v_{k,c}]= 0$. Otherwise, we consider the initial RB allocation is well performed and there is no need to update initial RB allocation and we set $\Omega^d_{c,k}=1$. 

\emph{Feedback for loop 2:}
We define RB usage  to evaluate the usage of RB $b_d$ allocated to  vO-DU $d$, where  RB usage $\tilde{\varphi}_{c,k}^d$ is given by:
\begin{equation}
	\label{eq:RBusage}
	\tilde{\varphi}_{c,k}^d=\frac{\sum_{k=1}^{K_d} b^{c,k}_d}{b_d}.
\end{equation} 

Based on RB usage and slice requirement satisfaction, we formulate the following optimization problem that maximizes resource utilization, while meeting resource constraints and QoS requirements in terms of latency:
\begin{subequations}
	\label{eq:problem_formulation0}
	\begin{align}
		&\underset{(\vect{y},\vect{z}, \vect{w})}{\text{max}}\ \  \sum \nolimits_{d  \in \mathcal{D}}   y^{c,d}_{b,k}\tilde{\varphi}_{c,k}^d + \sum \nolimits_{v  \in \mathcal{V}_k}w^{v,d}_{k,c}	\varphi^c_k
		\tag{\ref{eq:problem_formulation0}}\\
		& \text{subject to}\nonumber\\
		& \sum_{u\in \mathcal{V}_k}z^{v,m}_{t_i,f_i}\leq 1, \;  \forall m  \in \mathcal{M},\label{first:a}\\
		&\sum_{c\in \mathcal{C}}y^{c,d}_{b,k}\leq 1, \label{first:b}\\
		&	\sum_{v \in \mathcal{V}_k}^{}z^{v,m}_{t_i,f_i}\Omega^v_{c,k} R^v_i\leq b^{c,k}_d,\label{first:c}\\
		&	\sum_{v \in \mathcal{V}_k}^{}\lambda^v_{k,c}z^{v,m}_{t_i,f_i} o_c^{v,k}) \, \leq\varpi_{m,d}.\label{first:d}
	\end{align}
\end{subequations}
In the formulated optimization problem in (\ref{eq:problem_formulation0}), the constraint in (\ref{first:a}) ensures RB $(t^{c,k}_{i,d}, f^{c,k}_{i,d})$ can only be allocated to a single car. The constraint in (\ref{first:b}) guarantees that each slice $c$ associated to service $k$ is create at one vO-DU. The constraint in (\ref{first:c})  ensures that the RBs  allocated to cars ($R^v_i$ represents $(t^{c,k}_{i,d}, f^{c,k}_{i,d})$) do not exceed the available vO-DU resources. The constraint in (\ref{first:d}) is related to fronthaul network and it ensures that each node does not send more traffic than the fronthaul capacity. 

The problem in (\ref{eq:problem_formulation0}) is a combinatorial optimization problem, which is NP-hard and does not have an efficient polynomial-time solution. Also, an optimization problem that can lead to a stationary solution is not appropriate for resource auto-scaling because the resource auto-scaling process is a continuing, not stationary task \cite{lorido2014review}. Demands for network slices should be learned continuously to adapt to the change in workload and network environment. Therefore, we change (\ref{eq:problem_formulation0})  to a reward function so that it can reflect different QoS fulfillment,  workload changes, and network condition changes.

We formulate a reward function $	r_{t,c}(\vect{z}, \vect{w})$  for closed loop $1$ so that it can reflect intra-slice QoS fulfillment in terms of delay and  workload changes at time $t$:
\begin{multline}
	\label{eq:problem_formulation1}
	r_{t,c}(\vect{z}, \vect{w})=w^{v,d}_{k,c}	\varphi^c_k +	\Delta_m(\varpi_{m,d} -	\sum_{v \in \mathcal{V}_k}^{}\lambda^v_{k,c}z^{v,m}_{t_i,f_i} o_c^{v,k}) \\+ 	\Delta_v(1- \sum_{u\in \mathcal{V}_k}z^{v,m}_{t_i,f_i}) + \Delta_z(-\nu_c^d),
\end{multline}
where $ \nu_c^d=\sum_{u\in \mathcal{V}_k}z^{v,m}_{t_i,f_i}R^v_i\Omega^d_{c,k}-b^{c,k}_d$. 
We use $\Delta_m$ to denote the penalty of violating fronthaul resource constraint. $\Delta_v$ is penalty parameter for violating RB allocation constraint. $\Delta_z$ is the penalty parameter to ensure that intra-slice scaling does not violate the vO-DU RBs capacity constraint.

 We formulate a reward function $r_{t,d}(\vect{y})$ for closed loop $2$ to evaluate the RB $b_d$ utilization at vO-DU $d$ at time $t$:
\begin{multline}
	\label{eq:problem_formulation2}
	r_{t,d}(\vect{y})= y^{c,d}_{b,k}\tilde{\varphi}_{c,k}^d +  	\Delta_d(1-\sum_{c\in \mathcal{C}}y^{c,d}_{b,k})+ \\ \Delta_b( B - \sum_{c\in \mathcal{C}}y^{c,d}_{b,k}b^{c,k}_d+ \nu_c^d).
\end{multline}
where $\Delta_d$ is the penalty parameter to ensure each slice is managed by one vO-DU. We use $\Delta_b$ to denote the penalty that guarantees RB updates do not violate RB constraint. 

\emph{Connecting two loops:}
Closed loop 1 maximizes reward function $r_{t,c}(\vect{z}, \vect{w})$ by satisfying intra-slice QoS  in terms of delay and  workload changes at time $t$. On the other hand, closed loop $2$ needs to maximize reward $	r_{t,d}(\vect{y})$  and avoid violation of RB capacity constraints at vO-DU $d$. However, RB usage at vO-DU $d$ depends on intra-slice RB allocation. Therefore, $\nu_c^d$ enables to connect the actions of closed loop $1$ with actions of closed loop $2$. Closed loop $1$ needs to sends $\nu_c^d$ to closed loop $2$ as feedback that show the difference between RB demands  and allocated RBs to vO-DU.  Therefore,
we formulate a main reward function $r_t(\vect{y},\vect{z}, \vect{w})$ that interconnects the two proposed closed loops at time $t$, where $r_t(\vect{y},\vect{z}, \vect{w})$ is given by:
\begin{equation}
	\label{eq:problem_formulation3}
	r_t(\vect{y},\vect{z}, \vect{w})=	r_{t,d}(\vect{y}) + \phi_{dis} 	r_{t,c}(\vect{z}, \vect{w}).
\end{equation}
Since the closed loop two has to maximize reward in (\ref{eq:problem_formulation3}) that combines (\ref{eq:problem_formulation1}) and (\ref{eq:problem_formulation2}), where (\ref{eq:problem_formulation1}) is already maximized with closed loop one,  we introduce $\phi_{dis}$ as discount parameter for $r_{t,c}(\vect{z}, \vect{w})$ to allow the closed loop $2$ to put more emphasis on  (\ref{eq:problem_formulation2}).
 
\section{Proposed Solution} 
\label{sec:ProposedSolution}

In (\ref{eq:problem_formulation2}),  closed loop 2 at Near-RT RIC needs to deal with  actions $\mathcal{A}(\vect{y})$ consist of assigning initial RBs, keep initial RBs allocation ($\nu_c^d=0$), RBs scale-up ($\nu_c^d>0$), RBs scale-down ($\nu_c^d<0$) , and terminate RBs allocation for vO-DUs ($ \nu_c^d=-b^{c,k}_d$, i.e., $\sum_{u\in \mathcal{V}_k}z^{v,m}_{t_i,f_i}R^v_i\Omega^d_{c,k}=0$).  The states   $\mathcal{S}= \{(\vect{B}, \vect{D}, \vect{C})\}$ at Near-RT RIC consist of the states of RBs $\vect{B}$, vO-DUs $\vect{D}$, and slices $\vect{C}$ managed by vO-DUs.
On the other hand,  closed loop 1 needs to deal with actions $\mathcal{A}'(\vect{z}, \vect{w})$ consist of assigning initial RBs,  keep initial RBs allocation ($\Omega^d_{c,k}=1$), RBs scale-up ($\Psi^d_{c,k} = \beta^d_{c,k}$), RBs scale-down ($\Psi^d_{c,k} > \beta^d_{c,k}$), and terminate RBs allocation ($\Psi^d_{c,k} = \tilde{\beta}^d_{c,k}$) for cars. The states $\mathcal{S}'= \{( \vect{V}, \vect{\Omega},\vect{\Psi}) \}$ at RT-SC consist of  the states of $\vect{V}$ cars managed by slices, intra-slice orchestration $\vect{\Omega}$, and queue $\vect{\Psi}$.   The closed loop $1$ has direct access to the environment, observes cars' demands, and assigns RBs to cars. Based on queue status and intra-slice satisfaction, the closed loop $1$  can keep or update the RBs allocation for cars. Then, it gives feedback to closed loop $2$ so that closed loop $3$ can  have an overview of  $\mathcal{A}(\vect{y},\vect{z}, \vect{w})$, maximize (\ref{eq:problem_formulation2}),  and update RBs for vO-DUs $D$. Since the initial RBs allocation to the services of tenants is based on ARB, in RB auto-scaling  using $\nu_c^d$ and $\Omega^d_{c,k}$, we assume the InP and tenants can negotiate  flat rate increase or decrease on $J^{l, k*}_b(n^{l,k}_b)$.

\begin{figure}[t]
	\centering
	\includegraphics[width=0.85\columnwidth]{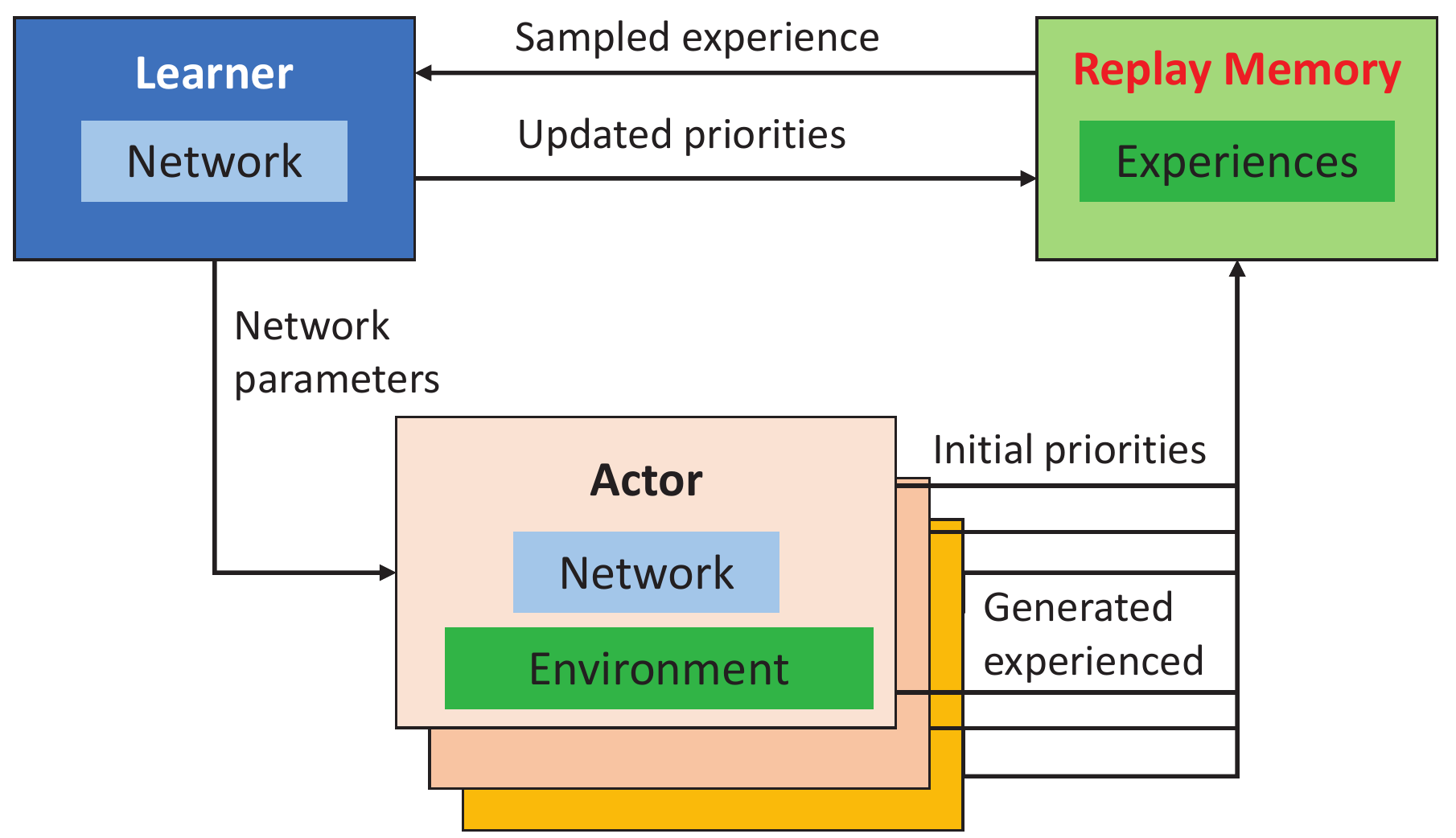}
	\caption{The Ape-X architecture \cite{horgan2018distributed}.}
	\label{fig:Apex}
\end{figure}

RL  or DRL \cite{kiran2021deep} can be applied to handle
the formulated rewards. However, finding one RL  or DRL model that uses two closed loops is a challenging issue. To overcome this issue, we choose Ape-X \cite{horgan2018distributed} shown in  Fig.	\ref{fig:Apex}  as distributed RL over other RL  or DRL approaches. 
\begin{figure*}[t]
	\centering
	\includegraphics[width=1.8\columnwidth]{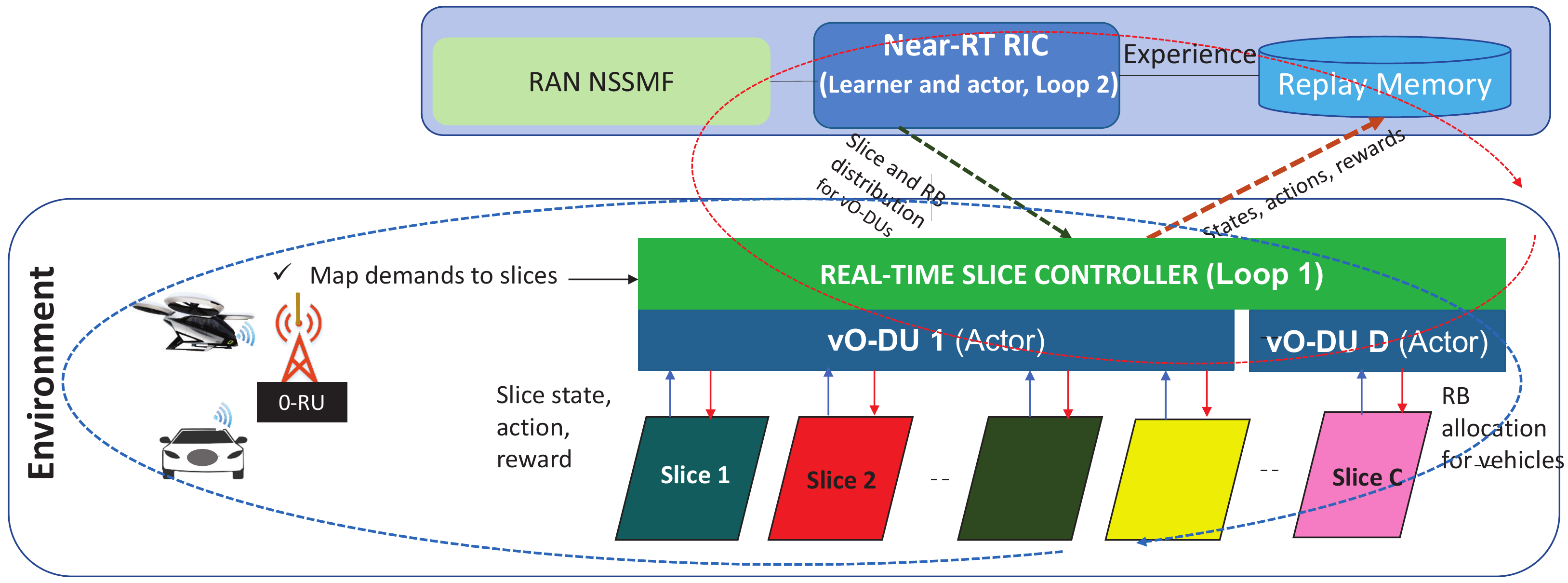}
	\caption{Interconnected closed loops in managing  RAN slicing resources serving cars.}
	\label{fig:SystemModelVehicleSlice}
\end{figure*}
Ape-X decomposes deep RL into two components. The first component interacts with the environment, implements,  and evaluates deep neural network. Then, it stores the observation data in a replay memory. We consider this process as acting, where the component is an actor. The second component samples batches of data from replay memory and updates the parameters. We consider this process as learning, where the second component is leaner. Ape-X can be combine with different learning algorithms, such as Deep Q Learning (DQN). In this work, we combined Ape-X with DQN \cite{fan2020theoretical}, where DQN integrates deep learning into Q-Learning.
The simplest form of  Q-Learning, which is called one-step Q-Learning, is given by:
\begin{equation}
	\begin{aligned}
		Q(s_t, a_t)= Q(s_t, a_t) + \alpha [r_{t+1} +\gamma_t Q(s_{t+1}, a) \\ - \; Q(s_{t}, a_t)],
	\end{aligned}
\end{equation}
where $\alpha$ is the learning rate and $a \in \mathcal{A}$ is an action that was taken in the state 
$s_t$ by an agent. $\gamma_t$  ($0 <\gamma_t\leq 1$) is discount factor. On the other hand, DQN uses standard feed-forward neural networks to calculate Q-Value. The DQN uses two networks,  Q-Network to  calculate Q-Value in the state $s_t$ and  target network to calculate Q-Value in the state $s_{t+1}$ such that:
\begin{equation}
	\begin{aligned}
		\label{eq:qdl}
		Q(s_t, a_t)= Q(s_t, a_t) + \alpha (r_{t+1} +\gamma_t\underset{\vect{a} }{\text{max}} Q(s_{t+1}, a) \\ - \; Q(s_{t}, a_t)).
	\end{aligned}
\end{equation}
The loss function $\Phi(\theta )$ to be minimized can be expressed as follows:
\begin{equation}
	\begin{aligned}
		\label{eq:loss2}
		\Phi_t(\theta)=\frac{1}{2}(\tilde{G}_t- Q(s_t, a_t,\theta))^2,
	\end{aligned}
\end{equation}
where $\theta$ represents parameters of the neural network and  $\tilde{G}_t$ is the return function. $\tilde{G}_t$ can be expressed as follows:
\begin{equation}
	\begin{aligned}
		\label{eq:problem_discounted rewards}
		\tilde{G}_t= r_{t+1} + \gamma r_{t+2}+ \dots + \gamma^{n-1} r_{t+n} + \\ \gamma^{n} Q(s_{t+n}, \underset{\vect{a} }{\text{argmax}} Q(s_{t+n}, a, \theta) , \theta^{-}).
	\end{aligned}
\end{equation}
In (\ref{eq:problem_discounted rewards}), $n$ is the number of steps. We use $t$ to represent a time index of sampling experience in replay memory. The experience sampling starts with state $s_t $, action $a_t$, and parameters of the target network $\theta^{-}$. We use $\mathcal{T}$ to denote the total number of time steps until the end of the training process.

Fig. \ref{fig:SystemModelVehicleSlice} shows the application of  Ape-X as solution to our problem. In our approach, Near-RT RIC acts as learner and actor  for closed loop $2$ and  vO-DUs acts as actors for closed loop $1$. In Algorithm \ref{algo:ODU-slicesRB}, Near-RT RIC initializes $\theta_0$ and $b^{c,k}_d$. Then, Near-RT RIC sends $b^{c,k}_d$ and $\theta_0$ to vO-DUs via RT-SC and save them to replay memory. Also, Algorithm \ref{algo:ODU-slicesRB}  keeps checking the replay memory to get updates from closed loop $1$ and computes the loss function $\Phi_t(\theta)$ and updates $\theta_t$ to $\theta_{t+1}$. Then, Near-RT RIC computes Temporal Difference (TD) error ($\gamma_t\underset{\vect{a} }{\text{max}} Q(s_{t+1}, a)  - \; Q(s_{t}, a_t)$) using DQN and updates replay memory and sends $\theta_{t+1}$ and updated RBs $b^{c,k}_d$  to RT-SC for vO-DUs.

\begin{algorithm}
	\caption{: RBs allocation to vO-DUs (Near-RT RIC as Learner and Actor).}
	\label{algo:ODU-slicesRB}
	\begin{algorithmic}[1]
		\STATE {\textbf{Input:} $\mathcal{T}$;}
		\STATE {Initialize $t = 0$;}	
		\STATE {$\theta_0 \gets$ InitializeLeaningParameter();}
		\STATE {$b^{c,k}_d \gets$ AssignRBtovODU();}
		\FORALL{$t=1$ to $(t=\mathcal{T})$}
		\STATE { $a_{t-1} \gets$ KeepUpdateSliceResourcetovODU(); } 
		\STATE {$\tilde{\varphi}_{c,k}^d \gets$ CalculatevODUutilization();} 
		\STATE {$r_{t,d}(\vect{y}) \gets$ CalculateReward();}
		\STATE {InLocalMemory.add(($s_{t-1}$, $a_{t-1}$, $r_{t,d}$,  $\gamma_t$));}
		\STATE { ${id}, \tau \gets$ GetSampleFromReplayMomory(); } 
		\STATE { $\Phi_t(\theta) \gets$ CalculateLoss($\tau; \theta_t$);} 
		\STATE { $\theta_{t+1}\gets$ UpdateLearningParameters($\Phi_t(\theta), \theta_t$);}
		\STATE {$b^{c,k}_d \gets$ UpdateRBAllocation();}
		\STATE {$r_t(\vect{y},\vect{z}, \vect{w}) \gets$ CalculateReward();}
		\STATE {$p \gets$ CalculateTD();}
		\STATE {InReplayMemory.SetTD(${id}, p, r_t$);} 
		\STATE {PeriodicallyUpdateReplayMemory()).} 
		\ENDFOR
	\end{algorithmic}
\end{algorithm}

In Algorithm \ref{algo:Intra-slicesRB}, vO-DU gets initial parameters from the learner and via  RT-SC  such as $\theta_0$  and  RBs $b^{c,k}_d$ and slices assigned to vO-DU. Then, vO-DU performs intra-slices actions. We use $\mathcal{T}'$ to denote the total number of time steps for vO-DU. Each vO-DU stores states, $\nu_c^d$, actions, rewards, and discount factors in local memory. In each period $\tilde{\mathcal{T}}$, states, orchestration parameters, actions, rewards, discount factors, and TD,  are sent to replay memory via RT-SC so that the Algorithm \ref{algo:ODU-slicesRB} can update $b^{c,k}_d$ and $\theta_0$. We assume that $\tilde{\mathcal{T}}$ is not the same for different vO-DUs.
\begin{algorithm}
	\caption{: Intra-slices RB allocation to cars (vO-DU as Actor).}
	\label{algo:Intra-slicesRB}
	%\Theta is the momory and T is time horizon
	\begin{algorithmic}[1]
		\STATE {\textbf{Input:} $\tilde{\mathcal{T}}, \mathcal{T}'$;}
		\STATE {Initialize $t = 0$;}	
		\STATE {$\theta_0 \gets$ GetLearningParameters();}
		\STATE {$b^{c,k}_d \gets$ vODUGetRBs();}
		\STATE {$s_0 \gets$ environment.initialize();}
		\FORALL{$t=1$ to $(t=\mathcal{T}')$}
		\STATE { $a'_{t-1} \gets$ KeepUpdateRBtoCars(); } 
		\STATE {$\varphi^c_k \gets$ CalculateNetworkSliceSatisfaction();}
		\STATE {$	\Omega^d_{c,k} \gets$ CalculateOrchestrationParameter();}
		\STATE {$		r_{t,c}(\vect{z}, \vect{w}) \gets$ CalculateReward();}
		\STATE {InLocalMemory.add(($s'_{t-1}$, $\Omega^d_{c,k}$, $\nu_c^d$, $a'_{t-1}$, $	r_{t,c}$,  $\gamma_t$));} 
		\IF{LocalBuffer$\geq \tilde{\mathcal{T}}$}
		\STATE {$\tau \gets$ LocalBuffer.Get($\tilde{\mathcal{T}}$);}
		\STATE {$p \gets$ CalculateTD($\tau$);} 
		\STATE {InReplayMemory.ADD($\tau, p$, $\Omega^d_{c,k}$, $\nu_c^d$, $s'_{t-1}$, $a'_{t-1}$, $r_{t,c}$, $\gamma_t$);} 
		\ENDIF
		\STATE {Periodically($\theta_t\gets$ GetLearningParameterUpdate());} 
		\STATE {Periodically($b^{c,k}_d\gets$ vODUGetRBsUpdate()).}
		\ENDFOR
	\end{algorithmic}
\end{algorithm}
\begin{theorem}Computational complexity of Algorithms \ref{algo:ODU-slicesRB} and \ref{algo:Intra-slicesRB} is $O(n)$.
	\label{RA-TPM1} 
\end{theorem}
\begin{proof} 
	In the Algorithm \ref{algo:ODU-slicesRB}, we have one loop at lines $5-18$, which depends on number of vO-DUs and slices. On the other hands, the Algorithm  \ref{algo:Intra-slicesRB} contains one loop at lines ($6-19$) and it depends on the number of vehicles. In extreme scenario, we may have $n$ number of vehicles, slices, and vO-DUs. As result, Algorithms \ref{algo:ODU-slicesRB} and   \ref{algo:Intra-slicesRB}  have computational complexity $O(n)$.
\end{proof} 
\begin{figure}[t]
	\centering
	\begin{minipage}{0.45\textwidth}
		\centering
		\includegraphics[width=0.95\columnwidth]{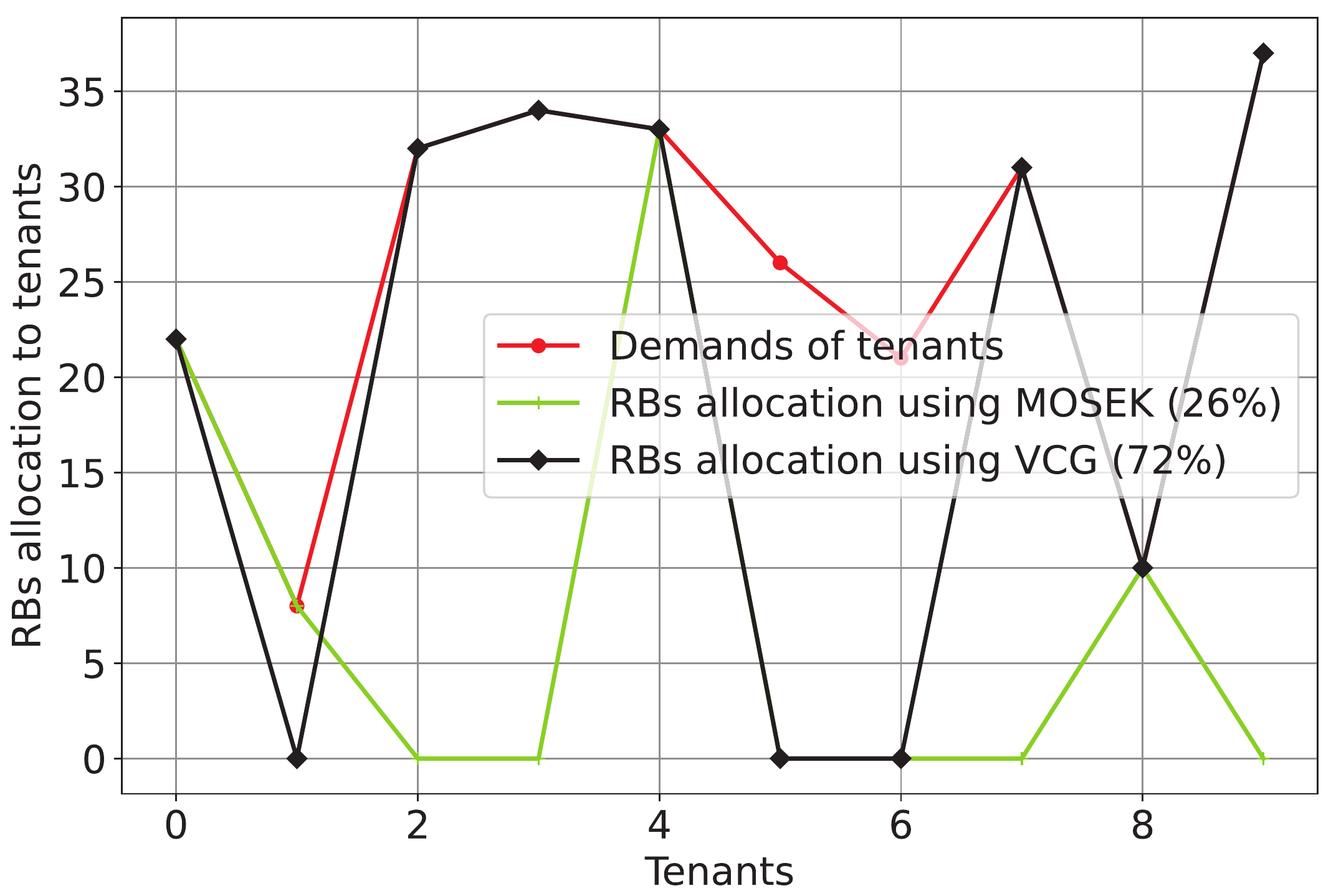}
		\caption{RBs allocation to tenants.}
		\label{fig:tenans}
	\end{minipage}
	\begin{minipage}{0.45\textwidth}
		\centering
		\includegraphics[width=0.95\columnwidth]{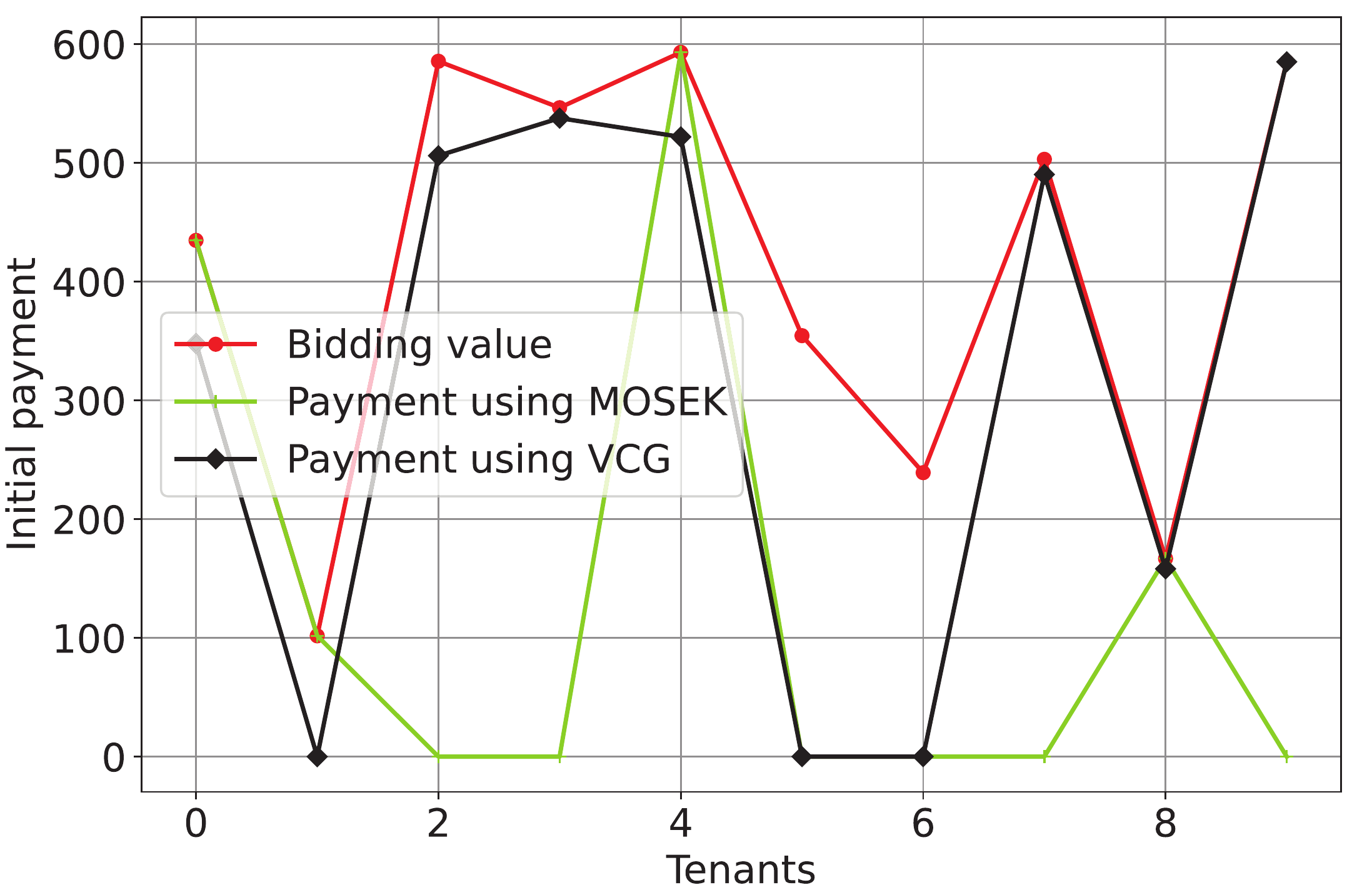}
		\caption{Tenant payments.}
		\label{fig:tenanspay}
	\end{minipage}
\end{figure}
\section{Performance Evaluation}
\label{sec:PerformanceEvaluation}
In this section, we present the performance evaluation of the
proposed closed loops for RAN slice resources management serving flying and ground-based cars. We use Python
\cite{nagpal2019python} for numerical analysis  and OpenAI Gym \cite{brockman2016openai} for making DRL environment.
\subsection{Simulation Setup}
\begin{figure}[t]
	\centering
		\begin{minipage}{0.45\textwidth}
		\centering
		\includegraphics[width=0.95\columnwidth]{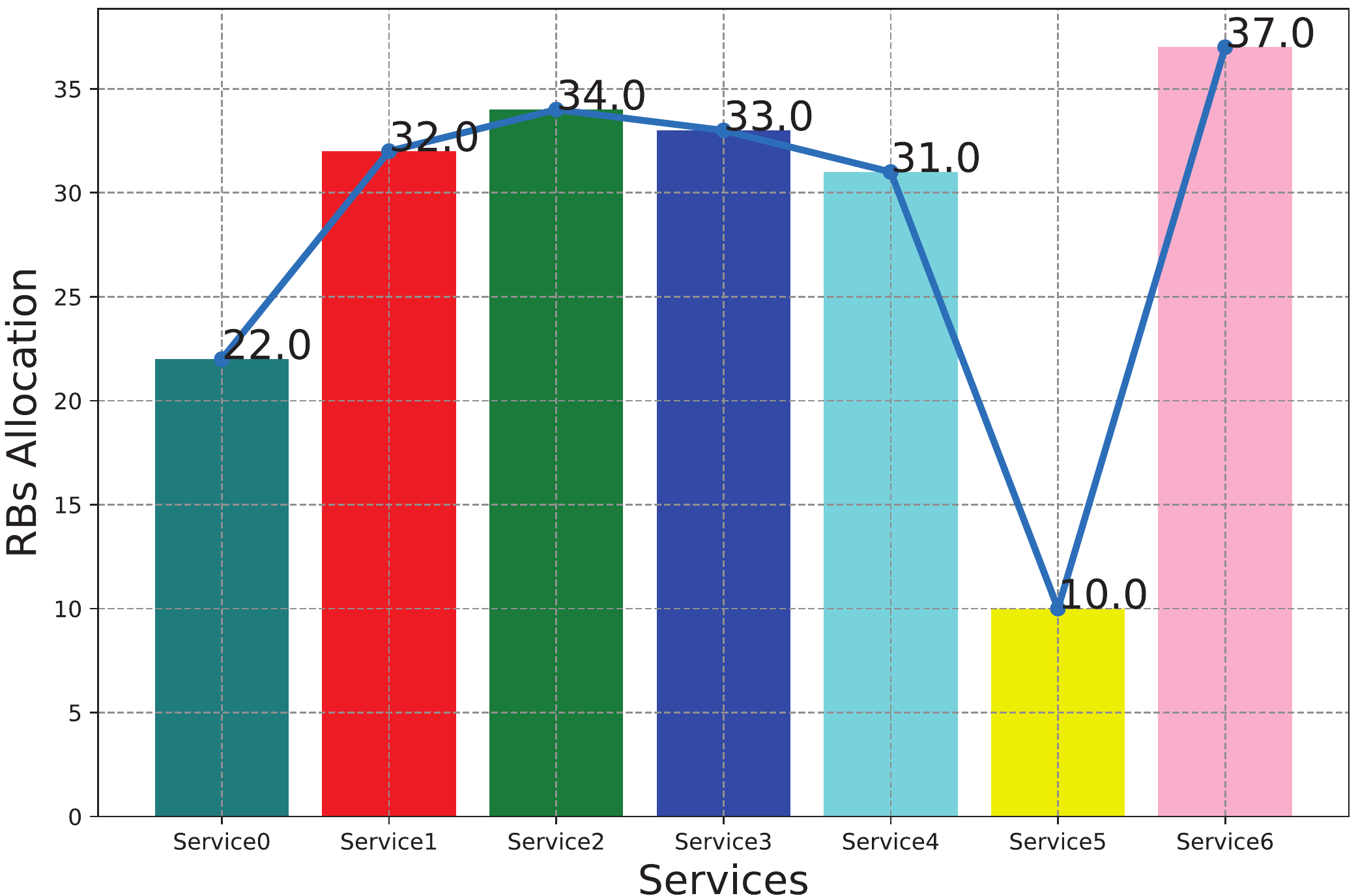}
		\caption{RBs allocation to services.}
		\label{fig:RB_services}
	\end{minipage}
	\begin{minipage}{0.45\textwidth}
		\centering
		\includegraphics[width=0.95\columnwidth]{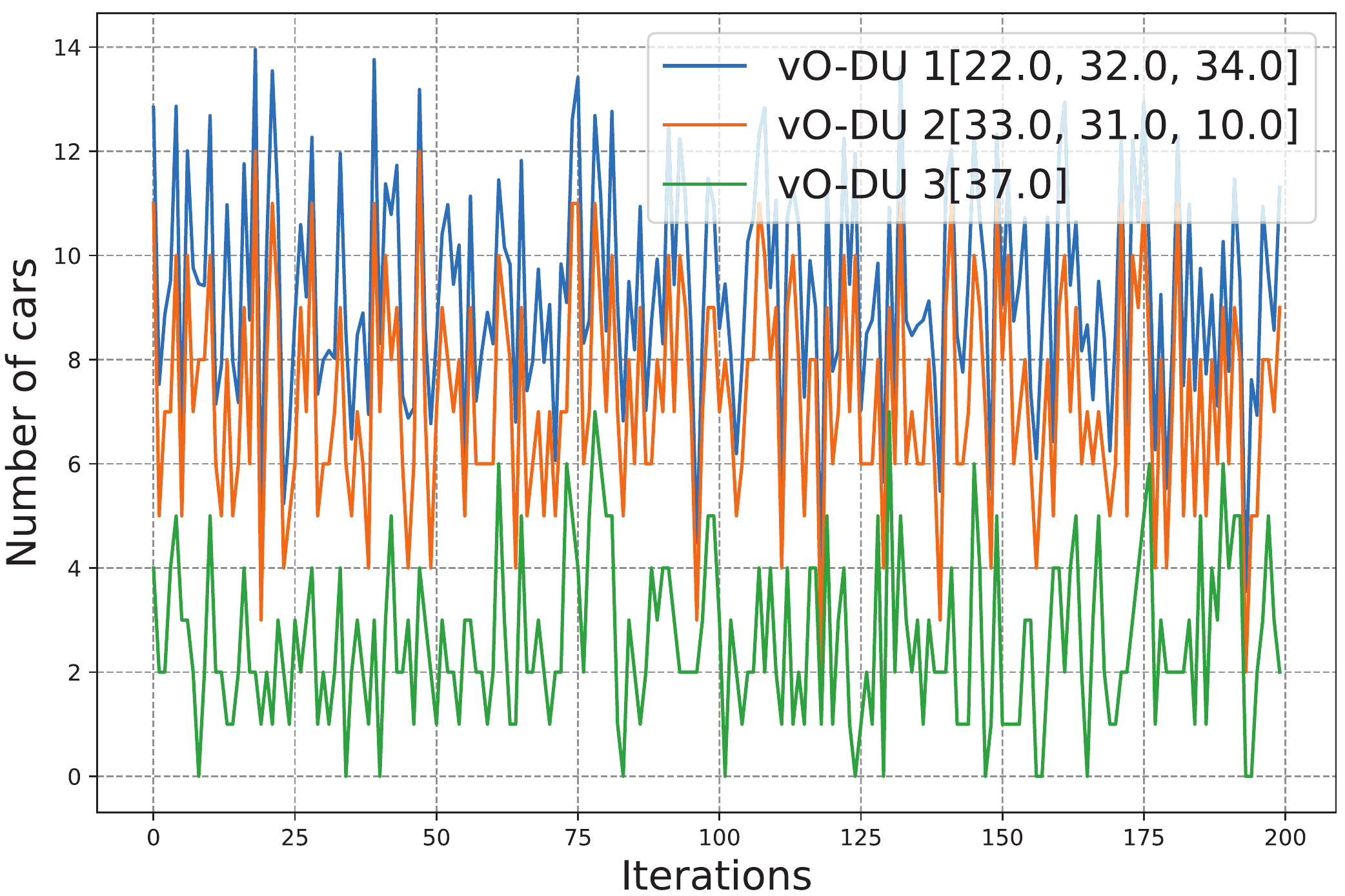}
		\caption{Number of cars per vO-DU .}
		\label{fig:vehicles_VDU}
	\end{minipage}
\end{figure}
  We use  $3$ flying cars and ground-based cars ranging from $10$ to $35$ cars. We use $6$ O-RUs and one edge cloud to provide a network connection to car. For the location of O-RUs, travel distances, time, and routes of flying and ground-based cars, we use  VeRoViz as a suite of tools designed for car routing from Optimator Lab at the University at Buffalo \cite{peng2020veroviz}. Since VeRoViz has drone features, we use drones as flying cars. We consider each car navigates/flies in the area of $6$ O-RUs. 
\begin{figure}[t]
		\centering
	\begin{minipage}{0.45\textwidth}
		\centering
		\includegraphics[width=0.95\columnwidth]{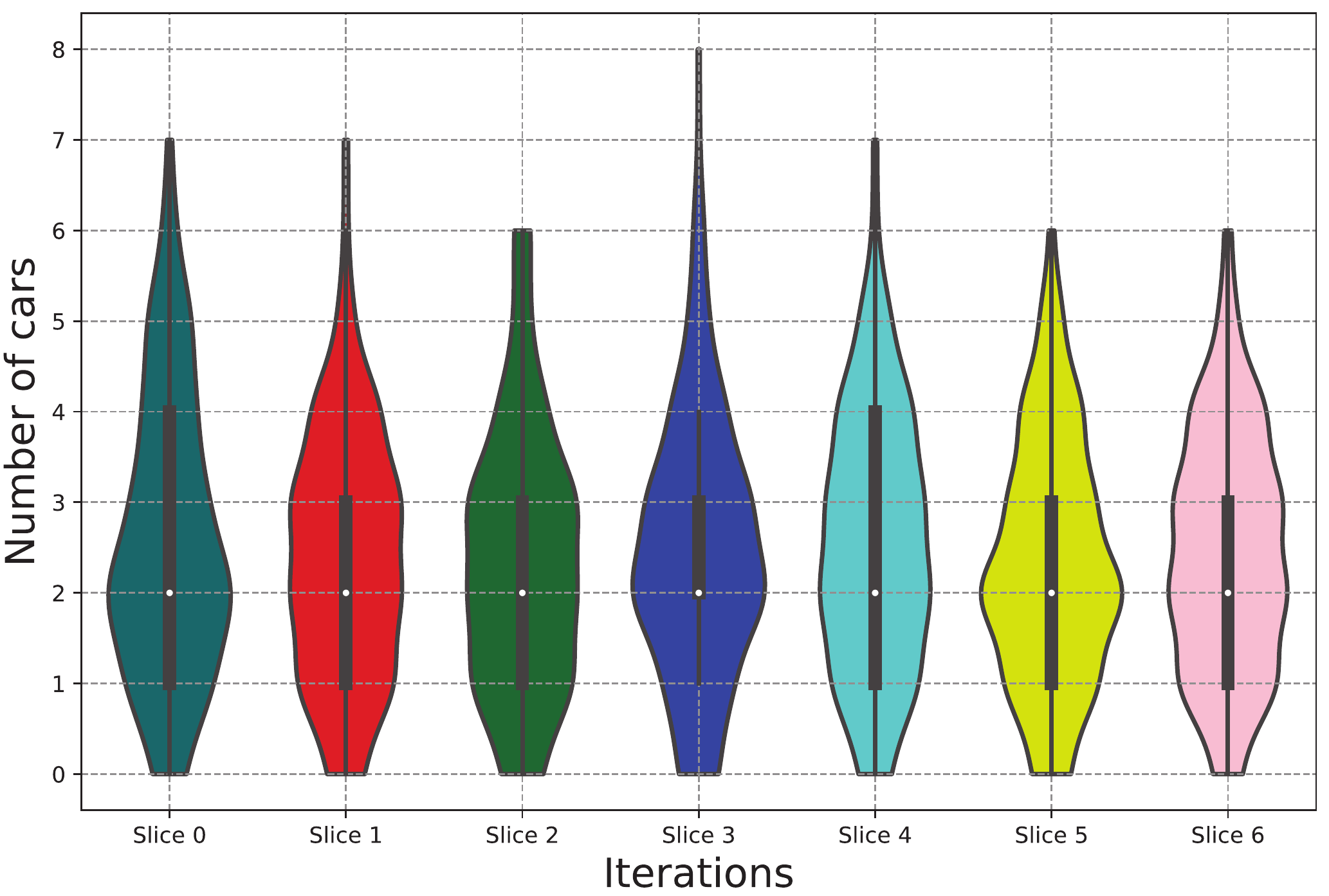}
		\caption{Number of cars per slice.}
		\label{fig:Slice_vehicles}
	\end{minipage}
	\begin{minipage}{0.45\textwidth}
		\centering
		\includegraphics[width=0.95\columnwidth]{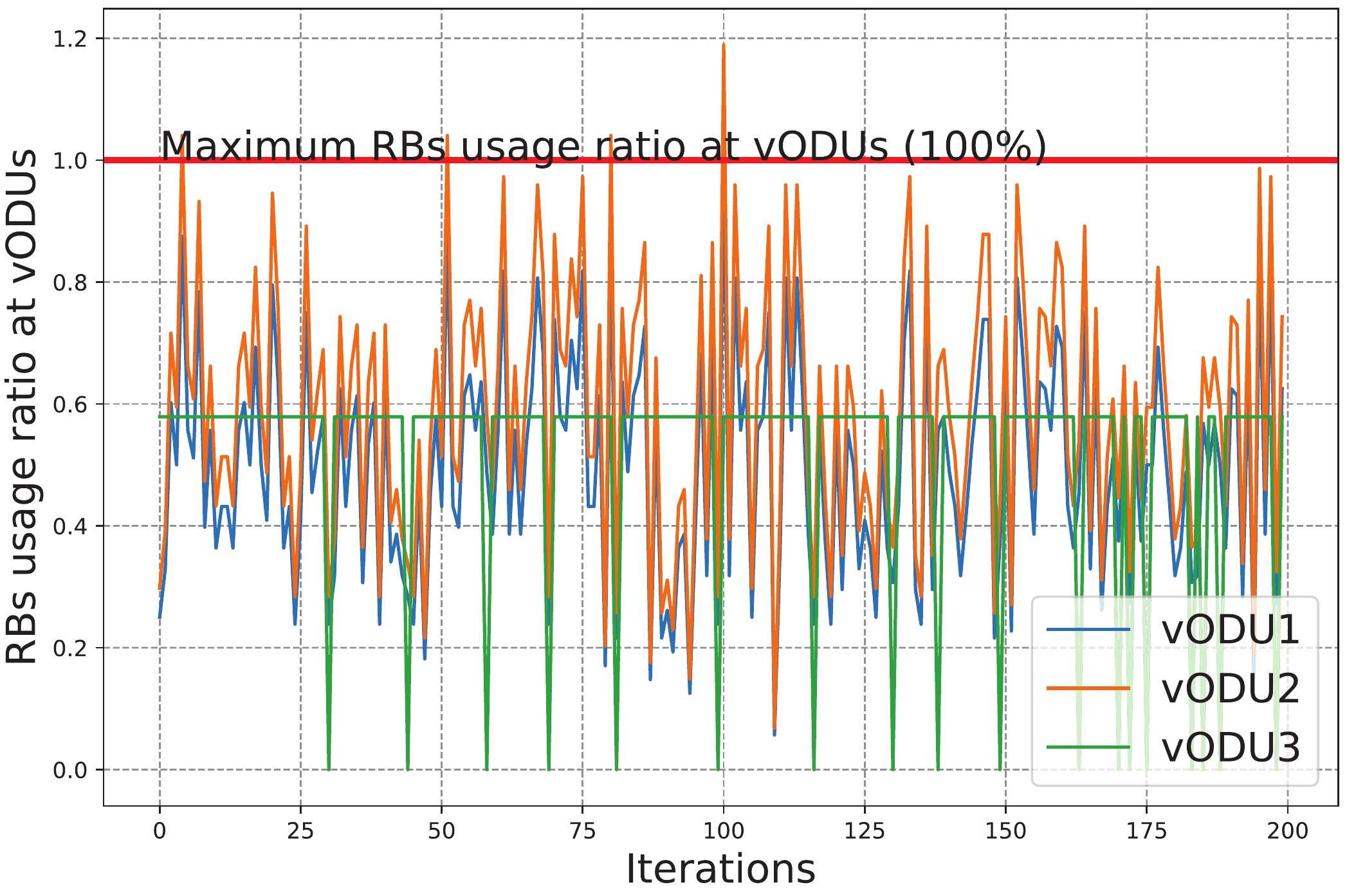}
		\caption{RBs usage ratio at vODUs.}
		\label{fig:Mean_evaluation_reward}
	\end{minipage}
\end{figure}
\begin{figure}[t]
		\centering
	\begin{minipage}{0.45\textwidth}
		\centering
		\includegraphics[width=0.95\textwidth]{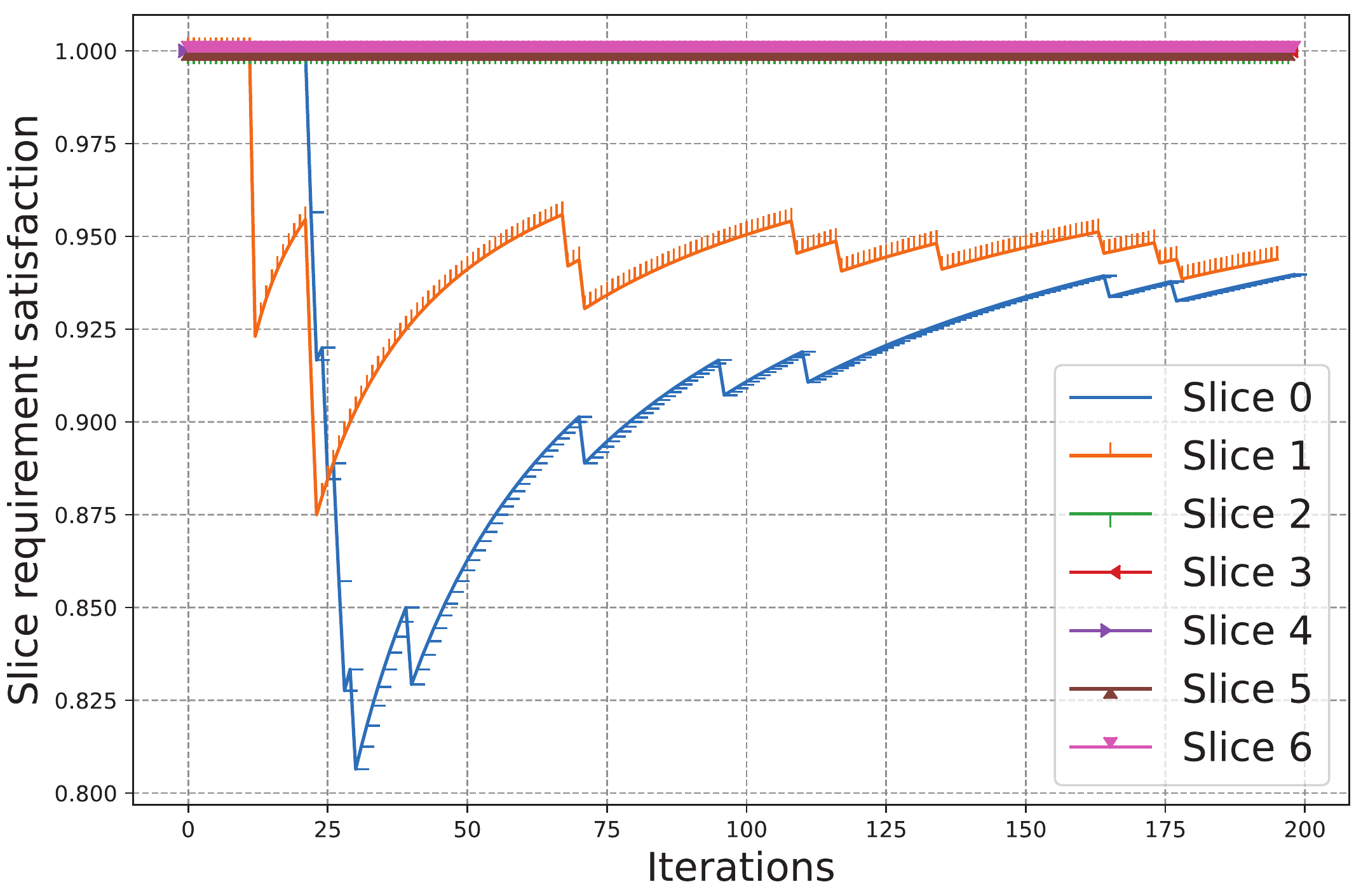}
		\caption{Network slice requirement satisfaction.}
		\label{fig:satisfaction}
	\end{minipage}
	\begin{minipage}{0.45\textwidth}
		\centering
		\includegraphics[width=0.95\textwidth]{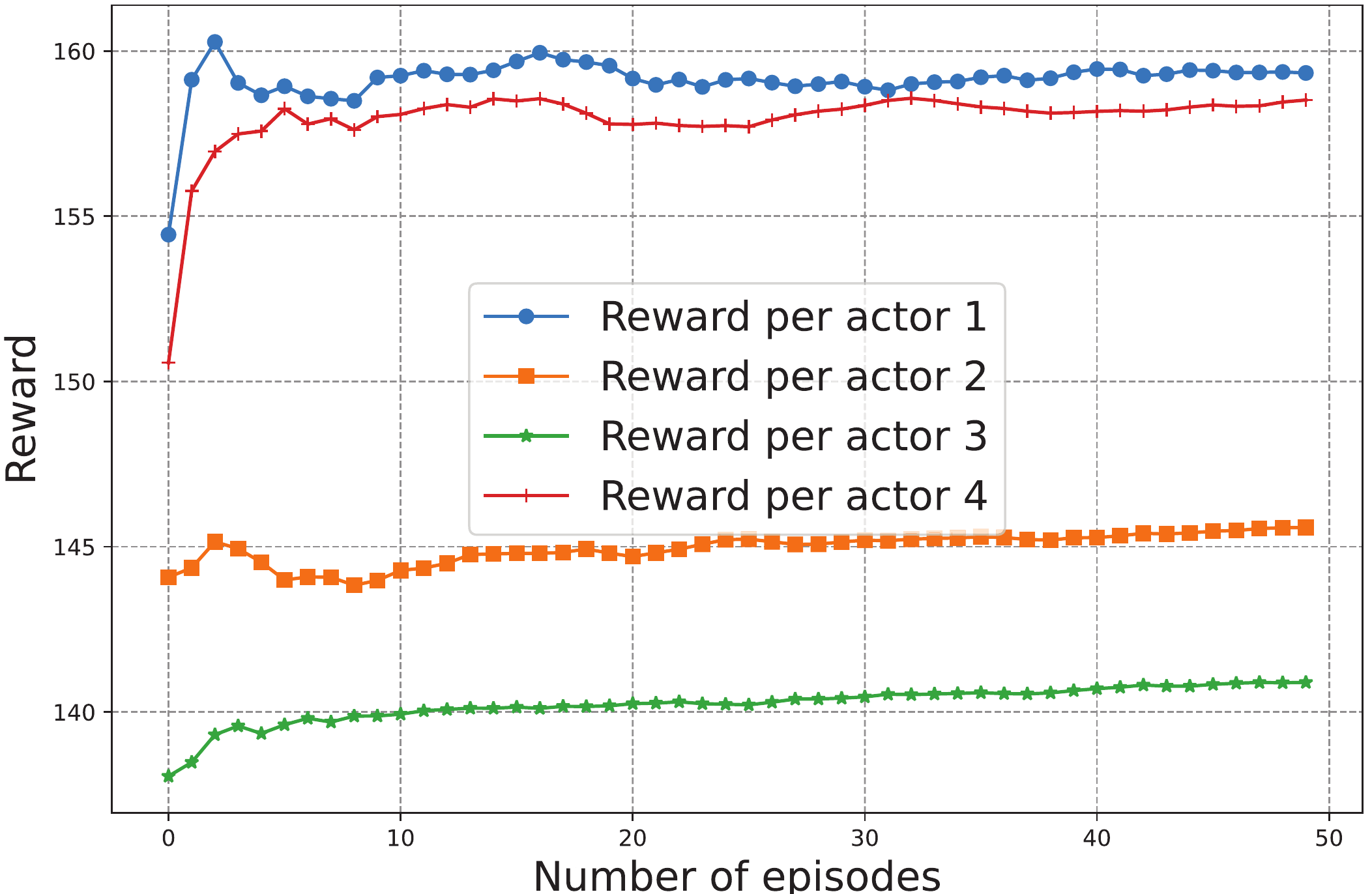}
		\caption{Reward maximization per actor.}
		\label{fig:reward1}
	\end{minipage}
\begin{minipage}{0.45\textwidth}
	\centering
	\includegraphics[width=0.95\textwidth]{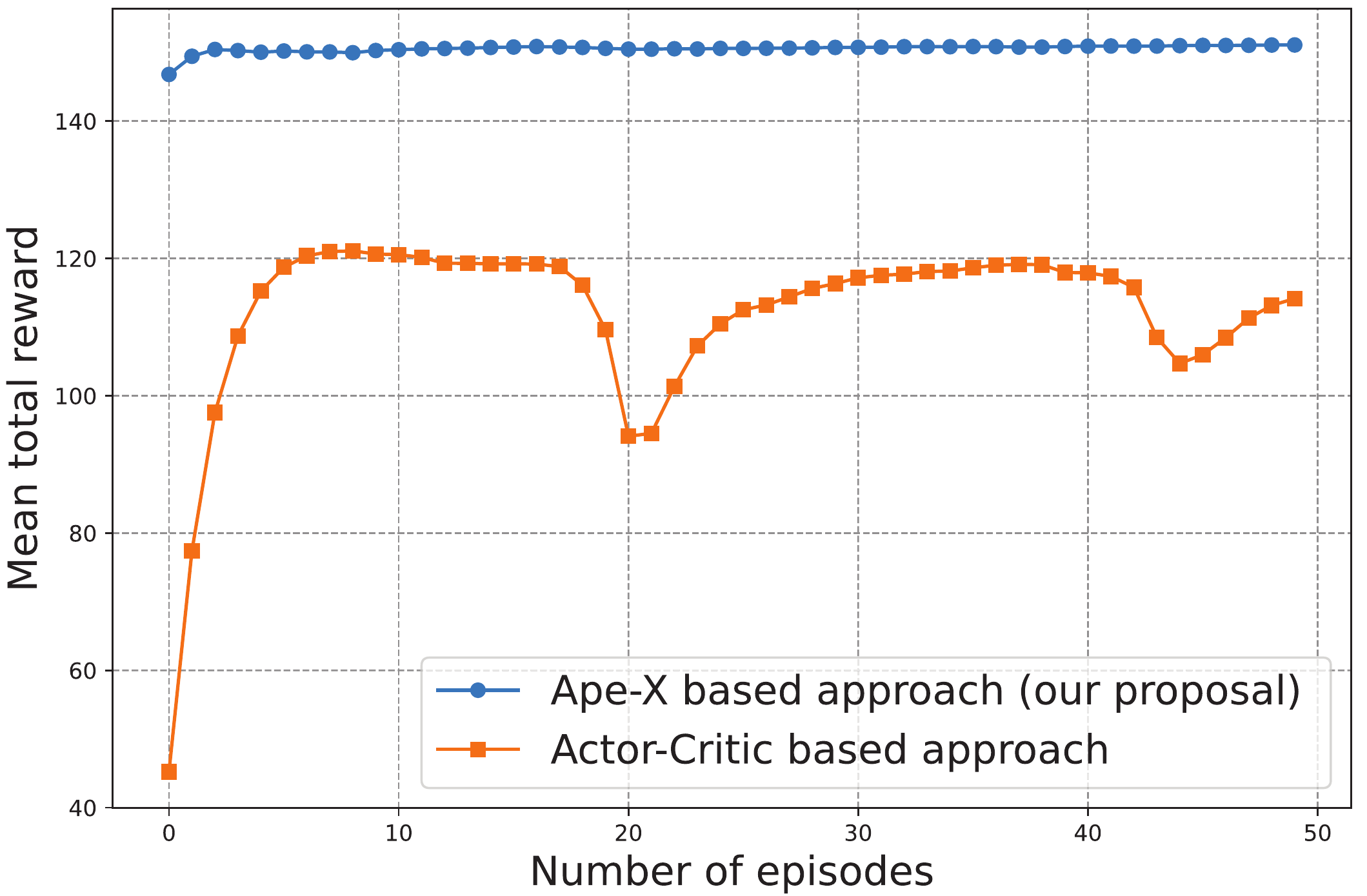}
	\caption{Mean total reward maximization.}
	\label{fig:Mean_evaluation_reward2}
\end{minipage}
\end{figure}
We use $100$ MHz channel bandwidth with $30$ kHz subcarrier spacing and $0.5$ millisecond TTI. The number of RBs is $273$ managed by $3$ vO-DUs, where each vO-DU initially has $b_d=91$ RBs. In ARB, we use $10$ tenants, where the demand $n^{l,k}_b$ of each tenant is in the range of  $6$ to $40$, and  $J^{l, k}_b$ is in the range from $10$ to $20$. We set $b_p=15$ and consider that the number of slices associated with services varies based on the output of the auction. We consider $7$ services from 5QI \cite{p1} such as advanced driving and remote driving, where the delay budget  $\tau^v_k$ is in the range from  $5$ to $300$ milliseconds. Each car chooses one or more service (s) randomly from the list of $7$ services. The packet size $o_c^{v,k}$ is generated  randomly in the range from $1$ kilobyte to $10$ megabytes.

As described  in \cite{MasteringReinforcement}, to implement Ape-X, we use Ray \cite{liang2017ray} and Keras with TensorFlow \cite{ramasubramanian2019deep}. In Ape-X,  for the neural network, we use the input layer of 3 neurons, two hidden layers of $64$ neurons per hidden layer, and an output layer of 4 neurons. The input of $3$ neurons corresponds to states. We assume initial RBs allocation can be performed based on initial demands. The four neurons in the output layer consist of $4$ actions: keep initial RBs allocation, RBs scale-up, RBs scale-down, and termination of RBs allocation. Time steps is set to $100000$, maximum sample size is set to $50000$ records, $\alpha = 0.0001$,  and $\gamma_t=0.99$. 

\subsection{Simulation Results}

The simulation results in Fig. \ref{fig:tenans} show RBs allocation to the tenants who provide services to the cars. Based on available RBs and bidding values ($J^{j, k}_b \geq b_p$),  $7$ tenants won the auction using the VCG and get $72$\% of the total RBs. Furthermore, we solve the optimization problem in (\ref{eq:sub11}) using MOSEK \cite{aps2017mosek} as mixed-integer optimization solver  and compare MOSEK solution with VCG solution. In MOSEK, only a small number of tenants of $J^{j, k}_b \geq b_p$ win the auction and get $26$\% of the total RBs. Even if we consider unallocated RBs as the residual resources that serve for RBs allocation scale-up, using MOSEK, InP remains with more unallocated RBs. Therefore, VCG has better performance than MOSEK.  The common behavior of VCG and MOSEK, they do not allow InP to allocate more than available RBs. Also,  as shown in Fig. \ref{fig:tenanspay}, with VCG and MOSEK, all winning tenants pay prices that are less or equal to their bidding values. In other words, our ARB satisfies individual rational and truthful bidding, where the winner pays a price that is less or equal to its bidding value, while the tenant who does not win ARB pays nothing. 

After the auction, hereafter, we use the results from VCG. Fig. \ref{fig:RB_services}  shows RBs allocation to the services of the tenants who won the auctions, where each service corresponds to one slice.  RBs of services are distributed to vO-DUs for scheduling purposes in the closed loop $2$. Fig. \ref{fig:vehicles_VDU} shows the RBs distributed to vO-DUs using the round-robin policy starting from vO-DU $1$, where vO-DU $1$ and vO-DU $2$ manages $3$ slices,  while vO-DU $3$ has one slice. Here, we remind that each vO-DU has $b_d=91$ RBs as the maximum limit, and RBs allocation to the slices at vO-DU has to respect RBs constraint ($\sum_{k=1}^{K_d} b^{c,k}_d \leq b_d$). In other words, the observation space of Ape-X for each vO-DU is in the range from $b_d= 0$ to $b_d=91$. RB allocation, scale-up, and scale-down should vary in this range.  In this figure, we show the number of cars getting service(s) from each vO-DU. Fig. \ref{fig:Slice_vehicles} shows the number of cars (minimum, first quartile, median, third quartile, and maximum) that use specific slices, where slice $3$ is more utilized than other slices.

Fig. \ref{fig:Mean_evaluation_reward} presents RBs usage ratio  defined in (\ref{eq:RBusage}) for vODUs. Since each vODU manages limited RBs, we consider $\tilde{\varphi}_{c,k}^d=1$ as the maximum RB usage ratio. In general, this figure shows that our approach satisfies vODUs resource constraints with a minor resource constraint violation at vO-DU $2$ (at $\tilde{\varphi}_{c,k}^d >1.0$ i.e., at more than $100\%$ utilization, the incoming request for RBs needs to be rejected). Furthermore, Fig \ref{fig:satisfaction}
shows  network slice requirement satisfaction in terms of delay as described in (\ref{eq:network_slice_requirement_satisfaction}), wherein most of the case our approach reaches $100\%$ slice requirement satisfaction except slices $0$ and $1$ managed by vO-DU $1$. 

Fig \ref{fig:reward1} presents the reward per actor using Ape-X. In other words, the rewards of  vODU 1 (actor $1$), vODU 2 (actor $2$), vODU 3 (actor $3$), and Near-RT RIC (actor 4). Here, we remind that vODUs focus on maximizing $r_{t,c}(\vect{z}, \vect{w})$ in closed loop $1$, while Near-RT RIC focuses on  maximizing $r_{t,d}(\vect{y})$ in closed loop $2$. Rewards are not the same for actors because different vODUs manage different slices. Also, the slices do not have the same numbers of RBs and serve varying numbers of vehicles. To compare Ape-X-based solution with other DRL approaches, we use reward function in  (\ref{eq:problem_formulation3}),  where discount parameter is set to $ \phi_{dis}=0.0018$. Fig. \ref{fig:Mean_evaluation_reward2} shows the mean of total reward using Ape-X and Actor-Critic DRL.  Actor-Critic is popular in DRL-based network slicing literature such as \cite{rezazadeh2021actor}. The results in this figure show that Ape-X has better performance than Actor-Critic DRL.

%Metrics
%%%%%%%%%%%%%%%%%%%%%%%%%%%%%%%
%Rutten, Eric, Nicolas Marchand, and Daniel Simon. "Feedback control as MAPE-K loop in autonomic computing." Software engineering for self-adaptive systems iii. assurances (2017): 349-373.
%deadline miss ratio, 
%rejection ratio,
% execution-time
%Computing A. An architectural blueprint for autonomic computing. IBM White Paper. 2006 Jun;31(2006):1-6.
% elapsed time to complete a process,
% percentage executed correctly
% the cost to execute a process
% throughput, 
% utilization
\section{Conclusion}
\label{sec:Conclusion}
This paper presented two-level closed loops for managing RAN slice resources serving flying and ground-based cars. We have used an auction mechanism for allocating RBs to the tenants who provide services to cars using slices. Then, we proposed two closed loops that complement each other, where closed loop $2$ distributes RBs to vO-DUs and closed loop $1$ at vO-DUs performs intra-slices RB scheduling for cars. Closed loop 1 sends resources utilization updates to closed loop $2$ so that the closed loop $2$ can update RBs distribution to vO-DUs. Using Ape-X as distributed reinforcement learning, the simulation results demonstrate that our approach satisfies more than 90\% vODUs resource constraints and network slice requirements. One of our future works is extending our framework with more performance evaluation in different simulation environments.
\bibliographystyle{IEEEtran}

\begin{IEEEbiography}[{\includegraphics[width=1in,height=1.25in,clip,keepaspectratio]{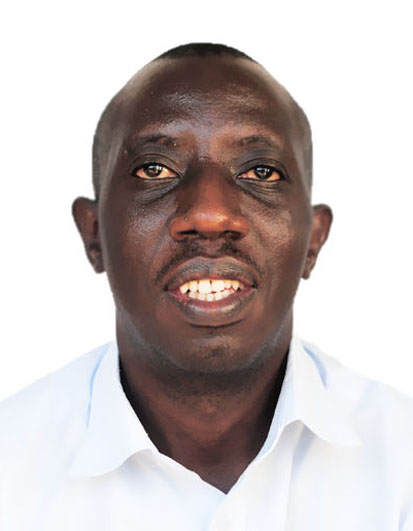}}]{Anselme Ndikumana} received  B.S. degree in Computer Science from the National University of Rwanda in 2007 and Ph.D. degree in Computer Engineering from Kyung Hee University, South Korea in August 2019. Since 2020, he has been	with the Synchromedia Lab, École de Technologie Supérieure, Université du Québec, Montréal, QC, Canada where he is currently a postdoctoral fellow. His professional experience includes Lecturer at the University of Lay Adventists of Kigali from 2019 to 2020, Chief Information System, a System Analyst, and a Database Administrator at Rwanda Utilities Regulatory Authority from 2008 to 2014. His research interest includes AI for wireless communication, multi-access edge computing, 5G networks, information-centric networking, and in-network caching.
\end{IEEEbiography}
\begin{IEEEbiography}[{\includegraphics[width=1in,height=1.25in,clip,keepaspectratio]{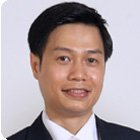}}]{Kim Khoa Nguyen} is Associate Professor in the Department of Electrical Engineering and the founder of the Laboratory on IoT and Cloud Computing at the University of Quebec’s  Ecole de technologie superieure. In the past, he served as CTO of Inocybe Technologies (now is Kontron Canada), a world’s leading company in software-defined networking (SDN) solutions. He was the architect of the Canarie’s GreenStar Network and led R\&D in large-scale projects with Ericsson, Ciena, Telus, InterDigital, and Ultra Electronics. He is the recipient of Microsoft Azure Global IoT Contest Award 2017, and Ciena’s Aspirational Prize 2018. He is the author of more than 100 publications, and holds several industrial patents. His expertise includes network optimization, cloud computing IoT, 5G, big data, machine learning, smart city, and high speed networks.
\end{IEEEbiography}
\begin{IEEEbiography}[{\includegraphics[width=1in,height=1.25in,clip,keepaspectratio]{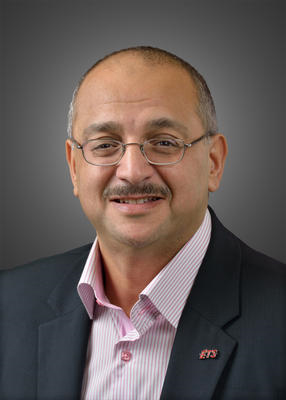}}]{Dr. Mohamed Cheriet} received his Bachelor, M.Sc. and Ph.D. degrees in Computer Science from USTHB (Algiers) and the University of Pierre \& Marie Curie (Paris VI) in 1984, 1985 and 1988 respectively. He was then a Postdoctoral Fellow at CNRS, Pont et Chaussées, Paris V, in 1988, and at CENPARMI, Concordia U., Montreal, in 1990. Since 1992, he has been a professor in the Systems Engineering department at the University of Quebec - École de Technologie Supérieure (ÉTS), Montreal, and was appointed full Professor there in 1998. Prof. Cheriet was the director of LIVIA Laboratory for Imagery, Vision, and Artificial Intelligence (2000-2006), and is the founder and director of Synchromedia Laboratory for multimedia communication in telepresence applications, since 1998.  Dr. Cheriet research has extensive experience in Sustainable and Intelligent Next Generation Systems. Dr. Cheriet is an expert in Computational Intelligence, Pattern Recognition, Machine Learning, Artificial Intelligence and Perception and their applications, more extensively in Networking and Image Processing. In addition, Dr. Cheriet has published more than 500 technical papers in the field and serves on the editorial boards of several renowned journals and international conferences. He held a Tier 1 Canada Research Chair on Sustainable and Smart Eco-Cloud (2013-2000), and lead the establishment of the first smart university campus in Canada, created as a hub for innovation and productivity at Montreal. Dr. Cheriet is the General Director of the FRQNT Strategic Cluster on the Operationalization of Sustainability Development, CIRODD (2019-2026). He is the Administrative Director of the \$12M CFI’2022 CEOS*Net Manufacturing Cloud Network. He is a 2016 Fellow of the International Association of Pattern Recognition (IAPR), a 2017 Fellow of the Canadian Academy of Engineering (CAE), a 2018 Fellow of the Engineering Institute of Canada (EIC), and a 2019 Fellow of Engineers Canada (EC). Dr. Cheriet is the recipient of the 2016 IEEE J.M. Ham Outstanding Engineering Educator Award, the 2013 ÉTS Research Excellence prize, for his outstanding contribution in green ICT, cloud computing, and big data analytics research areas, and the 2012 Queen Elizabeth II Diamond Jubilee Medal. He is a senior member of the IEEE, the founder and former Chair of the IEEE Montreal Chapter of Computational Intelligent Systems (CIS), a Steering Committee Member of the IEEE Sustainable ICT Initiative, and the Chair of ICT Emissions Working Group. He contributed 6 patents (3 granted), and the first standard ever, IEEE 1922.2, on real-time calculation of ICT emissions, in April 2020, with his IEEE Emissions Working Group.
\end{IEEEbiography}
\end{document}